\documentclass[12pt,preprint]{aastex}

\begin{document}

\title{Half-Megasecond {\em Chandra} Spectral Imaging of the
  Hot Circumgalactic Nebula around Quasar Mrk~231}

% \title{Half-Megasecond {\em Chandra} X-ray Spectral Imaging of the
%   Galactic Wind around the Nearest Quasar Mrk~231}

\author{S. Veilleux\altaffilmark{1,2}, S. H. Teng\altaffilmark{3},
D. S. N. Rupke\altaffilmark{4}, R. Maiolino\altaffilmark{5,6},
  \& E. Sturm\altaffilmark{7}}

\altaffiltext{1}{Department of Astronomy, University of Maryland,
  College Park, MD 20742, USA; veilleux@astro.umd.edu}

\altaffiltext{2}{Joint Space-Science Institute, University of Maryland,
  College Park, MD 20742, USA}

% \altaffiltext{3}{Astroparticle Physics Laboratory, NASA Goddard Space
%    Flight Center, Greenbelt, MD 20771, USA}
% 

\altaffiltext{3}{Observational Cosmology Laboratory, NASA Goddard Space
  Flight Center, Greenbelt, MD 20771, USA}

\altaffiltext{4}{Department of Physics, Rhodes College, Memphis, TN 38112, USA}

\altaffiltext{5}{Cavendish Laboratory, University of Cambridge, 19
  J.J. Thomson Ave., Cambridge, CB3 0HE, UK}

\altaffiltext{6}{Kavli Institute for Cosmology, Madingley Road,
  Cambridge, CB3 0HA, UK}

\altaffiltext{7}{Max-Planck-Institut f\"ur extraterrestrische
   Physik, Postfach 1312, D-85741 Garching, Germany}

\begin{abstract}
  A deep 400-ksec ACIS-S observation of the nearest quasar known,
  Mrk~231, is combined with archival 120-ksec data obtained with the
  same instrument and setup to carry out the first ever spatially
  resolved spectral analysis of a hot X-ray emitting circumgalactic
  nebula around a quasar.  The $\sim$65 $\times$ 50 kpc X-ray nebula
  shares no resemblance with the tidal debris seen at optical
  wavelengths. One notable exception is the small tidal arc $\sim$3.5
  kpc south of the nucleus where excess soft X-ray continuum emission
  and Si XIII 1.8 keV line emission are detected, consistent with star
  formation and its associated $\alpha$-element enhancement,
  respectively. An X-ray shadow is also detected at the location of
  the 15-kpc northern tidal tail, implying a foreground hydrogen
  column density of at least 2.5 $\times$ 10$^{21}$ cm$^{-2}$. The
  hard X-ray continuum emission within $\sim$6 kpc of the center is
  consistent with being due entirely to the bright central AGN and the
  wings of the {\em Chandra} point spread function. The soft X-ray
  spectrum of the outer ($>$6 kpc) portion of the nebula is best
  described as the sum of two thermal components of temperatures
  $\sim$3 and $\sim$8 million K with spatially uniform super-solar
  $\alpha$ element abundances, relative to iron. This result implies
  enhanced star formation activity over $\sim$10$^8$ yrs accompanied
  with redistribution of the metals on large scale.  The
  low-temperature thermal component is not present within 6 kpc of the
  nucleus, suggesting extra heating in this region from the
  circumnuclear starburst, the central quasar, or the wide-angle
  quasar-driven outflow identified from optical IFU spectroscopy on a
  scale of $\ga$3 kpc.  Significant azimuthal variations in the soft
  X-ray intensity are detected in the inner region where the outflow
  is present. The soft X-ray emission is weaker in the western
  quadrant, coincident with a deficit of H$\alpha$ emission and some
  of the largest columns of neutral gas outflowing from the
  nucleus. Shocks created by the interaction of the wind with the
  ambient ISM may heat the gas to high temperatures at this location.
  The tentative detection at the $\sim$2-sigma level of He-like Fe XXV
  6.7 keV line emission extending $\sim$3 kpc north-west of the
  nucleus provides some support to this scenario, if it is produced by
  $\sim$70 million K gas rather than high-mass X-ray binaries.
\end{abstract}

\keywords{galaxies: active --- galaxies: starburst --- ISM: jets and
  outflows --- quasars: individual (Mrk~231) --- X-rays: galaxies}

\section{Introduction}

In the past few years, Mrk~231 has become arguably the best laboratory
to study quasar feedback in action. One reason for this is the
proximity of Mrk~231: at a distance\footnote{Based on a redshift $z$ =
  0.0422 and a cosmology with $H_0$ = 73 km s$^{-1}$ Mpc$^{-1}$,
  $\Omega_{\rm matter}$ = 0.27, and $\Omega_{\rm vacuum}$ = 0.73.} of
only 178 Mpc, it is the nearest quasar known (Boksenberg et al. 1977)
and thus provides an excellent linear resolution (1$\arcsec$ = 0.863
kpc). But there are several other reasons why Mrk~231 has attracted
attention. It is one of the best local examples of powerful quasar and
starburst activity triggered by a recent merger (e.g., Hamilton \&
Keel 1987; Hutchings \& Neff 1987; Surace et al. 1998; Veilleux et
al. 2002, 2006). It is a morphologically disturbed ultraluminous
infrared galaxy (ULIRG) with an infrared (8 -- 1000 $\mu$m) luminosity
log[$L_{\rm IR}$/$L_\odot$] = 12.54 and a bolometric
luminosity$\footnote{This estimate for the bolometric luminosity
  includes all IR and non-IR contributions and is derived by simply
  assuming $L_{\rm BOL}$ = 1.15 L$_{\rm IR}$, typical for ULIRGs
  (e.g., Sanders \& Mirabel 1996).}$ log[$L_{\rm BOL}$/$L_\odot$]
$\approx$ 12.60 that is produced at $\sim$70\% by the quasar and
$\sim$30\% by a starburst with a star formation rate SFR $\sim$ 140
M$_\odot$ yr$^{-1}$ (Veilleux et al. 2009).  More relevant to the
issue of feedback, Mrk~231 is also a member of the rare class of iron
low-ionization broad absorption-line (FeLoBAL) quasars, with an
unresolved nuclear outflow with a velocity of up to $\sim$ --8000 km
s$^{-1}$ measured in Na~I~D 5890, 5896 \AA\ and several other optical
and ultraviolet lines (Boksenberg et al. 1977; Rudy, Foltz, \& Stocke
1985; Hutchings \& Neﬀ 1987; Boroson et al. 1991; Kollatschny,
Dietrich, \& Hagen 1992; Forster, Rich, \& McCarthy 1995; Smith et
al. 1995; Rupke, Veilleux, \& Sanders 2002; Gallagher et al. 2002,
2005; Veilleux et al. 2013a).  Finally, and most relevant to the
present paper, recent observations have revealed that Mrk~231 also
hosts a powerful spatially resolved optically-detected wind with
velocities in excess of $\sim$ --1000 km s$^{-1}$ (Rupke, Veilleux, \&
Sanders 2005c; Rupke \& Veilleux 2011, 2013, hereafter RV11 and RV13,
respectively).

The optical observations of RV11 and RV13 show that this wind is
predominantly neutral (traced in absorption by Na~I~D) and extends out
to at least $\sim$3 kpc. The neutral wind gas is symmetrically
distributed around the nucleus and presents a uniform velocity field
with no obvious spatial gradient. These results indicate the presence
of a wide-angle outflow that is expelling gas at a rate of at least
$\sim$200 M$_\odot$ yr$^{-1}$ out of the nuclear region (RV13). The
axis of this conical outflow is roughly along the minor axis of the
near face-on sub-kpc molecular / stellar disk (e.g., Downes \& Solomon
1998; Davies et al. 2004). Remarkably, a number of recent far-infrared
and mm-wave studies (Fischer et al. 2010; Feruglio et al. 2010; Sturm
et al. 2011; Aalto et al. 2012; Cicone et al. 2012; Gonzalez-Alfonso
et al. 2014) have inferred that a similar (if not larger) amount of
molecular gas is entrained in this outflow, reaching velocities that
are similar to those measured optically in the spatially resolved
neutral wind. This multi-phase wind is thus potentially a
life-changing event for Mrk~231, capable of displacing most of the
neutral and molecular gas from the nucleus, and thus quenching nuclear
star formation and also possibly the quasar activity, on a timescale
of only a few $\times$ 10$^6$ yr.  This is the essence of quasar
feedback, which is purported to transform many gas-rich mergers into
inactive bright elliptical galaxies (di Matteo, Springel, \& Hernquist
2005; Murray, Quataert, \& Thompson 2005; Veilleux, Cecil, \&
Bland-Hawthorn 2005; Veilleux et al. 2013b; Cicone et al. 2014).

Not surprisingly, Mrk~231 has been the subject of numerous X-ray
studies, using the whole gamut of high-energy space facilities,
starting with {\em Einstein}, {\em ROSAT}, {\em ASCA}, and {\em CGRO}
(e.g., Eales \& Arnaud 1988; Dermer et al. 1997; Turner 1999; Maloney
\& Reynolds 2000).  It was observed with {\em Chandra} on multiple
occasions: once in 2000 ($\sim$40~ksec; Gallagher et al.\ 2002), three
times in 2003 ($\sim$40~ksec each spanning over three weeks; Gallagher
et al. 2005), and once again in 2010 ($\sim$5-ksec snapshot; PI
Garmire). The 2000-2003 data were independently re-analyzed by Ptak et
al.\ (2003), Grimes et al.\ (2005), and Teng \& Veilleux
(2010). Mrk~231 was observed with {\em XMM-Newton} in 2001
($\sim$20~ksec; Turner \& Kraemer 2003) and at 12 $-$ 60 keV with {\em
  Beppo}SAX at the end of 2001 (Braito et al.\ 2004).  The combined
{\em XMM-Newton} + {\em Beppo}SAX spectrum of the nucleus suggests the
existence of a highly absorbed ($N_H \sim 2 \times
10^{24}$~cm$^{-2}$), powerful (intrinsic 2-10 keV luminosity of $\sim1
\times 10^{44}$ erg s$^{-1}$) AGN. The direct AGN continuum appears to
be only detected beyond $\sim$10 keV, while reprocessed radiation
through scattering and/or reflection dominates the observed 2-10 keV
emission (Braito et al. 2004).  A more recent (April 2011) 200-ksec
observation of Mrk~231 with {\em Suzaku} by Piconcelli et al. (2013)
indicates that the absorbing material is patchy, now letting through a
significantly larger fraction of the direct $<$10 keV AGN continuum
than in 2001. This patchy material may also be responsible for the
scattering/reflection itself and the optical/UV LoBAL systems (e.g.,
Rupke et al.\ 2002, 2005c; Gallagher et al. 2005; Veilleux et
al. 2013a and references therein). The recent high-energy (3 -- 30
keV) {\em NuSTAR} observation of Mrk~231, supplemented with our new
and simultaneous low-energy (0.5-8 keV) data from {\em Chandra}, puts
into question the existence of Compton-thick material in front of this
source (Teng et al. 2014). Mrk~231 was detected at high energies,
though at much fainter flux levels than previously reported, likely
due to contamination in the large apertures of previous non-focusing
hard X-ray telescopes. The full band (0.5 -- 30 keV) X-ray spectrum of
Mrk~231 suggests that the AGN in this object is absorbed by a patchy
and Compton-thin column of $N_{\rm H,~AGN}$ = $1.12^{+0.29}_{-0.24}
\times 10^{23}$ cm$^{-2}$.

A strong thermal (Mewe-Kaastra-Liedahl or MEKAL) component with $kT
\sim 0.7 - 0.8$ keV is also contributing to the soft X-ray band of
the nuclear spectrum (e.g., Gallagher et al. 2002, 2005; Ptak et
al. 2003; Teng \& Veilleux 2010; Teng et al. 2014).  But this nuclear
soft X-ray emission is just the tip of the iceberg: The {\em Chandra}
ACIS-S observations by Gallagher et al. (2002, 2005) have revealed a
spectacular soft X-ray nebula extending out to $\sim$30$\arcsec$
($\sim$25 kpc) from the nucleus of Mrk~231 (see also Grimes et
al. 2005).  A reanalysis of these data reveals that the morphology of
the inner portion of this nebula is similar to that of the H$\alpha$
emission mapped by RV11. In particular, there appear to be soft X-ray
enhancements immediately east of the nucleus and $\sim$3.5~kpc south
of the nucleus, respectively, coincident with an H~II region and the
tidal arc seen in RV11. These data show tantalizing evidence for
spatial variations in the properties of the X-ray gas, but the counts
are not sufficient to draw statistically significant conclusions.

The present paper describes the results from our analysis of a
considerably deeper {\em Chandra} ACIS-S data set, combining a new
400-ksec observation with the 2003 archival data obtained with the
same instrument and in the same configuration.
The results from a detailed analysis of the new {\em Chandra} data on
the central AGN, in combination with the high-energy {\em NuSTAR}
spectrum, were presented in Teng et al. (2014).  Here, we focus our
attention on the extended X-ray emission of Mrk~231 and discuss
whether some of this emission may relate to the spatially resolved
galactic wind.  The acquisition of the data is discussed in \S 2,
followed by a description of the results from the image and spectral
analyses (\S 3) and a discussion of the implications (\S 4). The
conclusions are summarized in \S 5.

\section{Observations}

The rationale behind the setup used for the new 400-ksec observation
(PID 13700587; PI Veilleux) was to match the observational parameters
of the 3 $\times$ 40-ksec exposures obtained in 2003 by Gallagher et
al. (2005) and thus facilitate the task of combining both data sets
into a single 0.52-Msec exposure. The other {\em Chandra} data sets
obtained before or after 2003 are considerably shallower or acquired
in a different mode so no attempt was made to combine them with the
present data set. Mrk~231 was aimed at the back-illuminated S3
detector of ACIS (Garmire et al. 2003). Due to scheduling constraints,
the planned 400-ksec observation was divided into three segments of
40.97, 182.02, and 177.99 ksec, obtained close in time (23, 24, and 27
August 2012, respectively) to reduce the effects of variability in the
AGN and background (e.g., solar flares). All of the observations were
performed in 1/2 subarray mode in order to avoid pileup and take
advantage of {\em Chandra}'s excellent angular resolution.

\section{Results}

\subsection{Image Analysis}

The data from the 2003 and 2012 epochs were combined together. Both
epochs were reduced using CIAO version 4.5 and CALDB version
4.5.6. This processing incorporates energy-dependent subpixel event
repositioning (EDSER algorithm; Li et al. 2004) which delivers
optimized image resolution by subpixel repositioning of individual
X-ray events. These data were re-processed to verify that this
optimization was indeed carried out properly and to check against
flares using the {\em Python}-based ``deflare'' routine, which
discards any data with count rate that is 3 standard deviations from
the mean of the distribution. We make no attempt to deconvolve the
data using, e.g., Lucy or EMC2 algorithms (Lucy 1974; Esch et
al. 2004; Karovska et al. 2005). This stategy better preserves diffuse features
and possible slight asymmetries in the point spread function
(PSF).\footnote{CIAO version 4.5 manual:
  http://cxc.harvard.edu/ciao/caveats/psf\_artifact.html.} The image
data were merged for analysis using the CIAO script ``merge$\_$obs''.
All of the data were reprojected to a common reference position.  This
minimizes the effect of relative astrometric errors on the final merged
image which might produce false morphological features.  The absolute
astrometric error for ACIS-S is $\sim$0$\farcs$6.  We checked our final
image against the VLA FIRST position of Mrk 231.  The absolute
astrometric offset was measured to be 0$\farcs$09.

The images derived from the combined data set are shown in Figure 1.
The full-band (0.5 -- 8 keV) images (top row in Figure 1) show a large
complex of emission extending over a total scale of at least
80\arcsec\ or $\sim$65 kpc in the north-south direction and 60\arcsec\
or $\sim$50 kpc in the east-west direction. This emission is not
distributed symmetrically with respect to the nucleus, extending
further to the south ($\sim$40 kpc) than to the north ($\sim$18 kpc),
and further east ($\sim$26 kpc) than west ($\sim$16 kpc). This
high-S/N data set also reveals that the X-ray emission is highly
clumpy or filamentary. The morphology of the large-scale X-ray
emission does not share a strong resemblance with that of the tidal
debris (upper middle panel of Figure 1; RV11; Koda et al. 2009). In
fact, the X-ray emission on the north side is {\em weaker} at the
position of the tidal tail.\footnote{The X-ray emission at the
  position of the tidal tail that extends more than 30 kpc south of
  the nucleus (Koda et al. 2009) also seems weaker, but the effect is
  less significant (see top middle panel of Figure 1). }  This
indicates that foreground tidal material absorbs some of the X-ray
photons along the line of sight. Iwasawa et al. (2011) reported a
similar X-ray ``shadow'' of a tidal tail in the ULIRG Mrk 273.  In \S
3.2.4, we derive a lower limit on the column density of the northern
tidal tail in Mrk~231 based on the apparent deficit of X-ray emission
at this location.

However, it is also clear that not all tidal features cast an X-ray
shadow. The bright tidal arc located $\sim$3.5 kpc south of the
nucleus coincides with X-ray enhancements (upper right panel of Figure
1). This tidal feature is known to be forming stars and may even be
powering its own supernova-driven outflow (RV11). The spectral
properties of the X-ray emitting gas at the location of this tidal arc
are discussed in \S 3.2.3; the results from this spectral analysis
confirm the presence of star-forming complexes at this location.

The false color maps shown on the bottom row of Figure 1 indicate a
trend with distance from the nucleus where the extended emission is
distinctly softer than the nuclear and circumnuclear emission. This is
illustrated more clearly in Figures 2 and 3, where the
azimuthally-averaged radial profile and two-dimensional map of the
hardness ratio [defined as ($HX - SX$) / ($HX + SX$), where HX and SX
are the 2 -- 8 keV and 0.5 -- 2 keV fluxes, respectively] are
presented. Also shown in Figure 2 are the azimuthally-averaged radial
profiles of the background-subtracted soft and hard X-ray emission fit
with a two-component $\beta$-model:
$$\Sigma(R) = \Sigma_0 [1 + (R/R_0)^2]^{-3\beta + 0.5},$$
where $\Sigma(R)$ is the azimuthally-averaged surface brightness
profile, $\Sigma_0$ is the central surface brightness, $R_0$ is the
core radius, and $\beta$ is the power-law index that quantifies the
slope at large radii. Note the dependence of the surface brightness
profile on X-ray energies, resulting in variations of the hardness
ratio. While the hardness ratio profile within $\sim$2 kpc is somewhat
affected by the energy dependence of the central PSF associated with
the AGN (the PSF is broader at higher energies; Wang et al. 2014), the
small values of the hardness ratios at $R \ga$ 2 kpc, where the flux
contribution from the central PSF is considerably smaller, reflect the
softness of the extended nebula. The right panel in Figure 3 also
suggests the presence of a slight asymmetry in the hardness ratio map
of the central region ($R \la$ 2 kpc), with higher hardness ratios
measured west of the nucleus. This result will be revisited in \S 3.2,
taking into account the spectral responses of {\em Chandra} at various
epochs.

\subsection{Spectral Analysis}

For each observation, individual spectra were extracted from
pre-determined extraction regions (see Figure 4) and then combined
together into a merged spectrum for each region using the ``combine''
keyword in the {\em specextract} function in CIAO 4.5. All spectra
used the same background extraction region to assure that differences
are not due to changes in the background since we expect the
variability in the background to be small from observation to
observation.  All spectra were binned to at least 15 counts per bin,
unless noted otherwise (e.g., when the counts were high, we binned by
S/N). XSPEC version 12.8.0k was used for this analysis. The errors
quoted below are at the 90\% confidence level.

The different regions were modeled {\em simultaneously} using a
procedure that is similar to that used to fit the quasars in Teng \&
Veilleux (2010). We initially started with the nuclear portion of the
{\em NuSTAR} model (hereafter called the nuclear {\em NuSTAR} model)
i.e., the model that was found by Teng et al. (2014) to best fit the
combined {\em NuSTAR} + {\em Chandra} nuclear data. In this model, the
direct AGN emission is absorbed and scattered by a patchy torus with
$N_{\rm H,~AGN}$ = $1.12^{+0.29}_{-0.24} \times 10^{23}$
cm$^{-2}$. More specifically, the best-fit model includes a leaky
MYTorus component for the AGN emission (Murphy \& Yaqoob 2009), an Fe
line at 6.7 keV from He-like Fe~XXV, and a heavily obscured ($N_{\rm
  H,~nuclear~HMXB}$ = 2.4 $\times$ 10$^{23}$ cm$^{-2}$) component from
high-mass X-ray binaries (HMXBs) associated with the nuclear star
formation (SFR of $\sim$140 $M_\odot$ yr$^{-1}$). The {\em NuSTAR}
model of Teng et al. (2014; see Table 1 in that paper for an equation
form of the best-fit {\em NuSTAR} model) also includes a two-component
MEKAL (Mewe-Kaastra-Liedahl) model of the emission from the hot diffuse
gas within the NuSTAR aperture (recall that the half-power diameter of
NuSTAR is $\sim$58$\arcsec$). Since our {\em Chandra} observations
resolve this diffuse emission, these MEKAL components are treated
separately from the nuclear component in the present paper (their
intensities and temperatures are allowed to vary independently of the
nuclear {\em NuSTAR} model; see below).

The flux of the nuclear {\em NuSTAR} model was allowed to vary with
position in the nebula, but the relative intensities of the various
components of this model were held fixed.  Previous studies (e.g.,
Gallagher et al. 2002; Teng et al. 2014) have shown that the hot
diffuse gas is best described as the sum of up to two thermal (MEKAL)
plasma components. For completeness, we also explore in this paper the
possibility that the hot diffuse gas is shocked plasma.  We confirm
that the best-fit models of our deeper {\em Chandra} data requires at
least one MEKAL component or one shock component to explain the soft
X-ray emission. In all fits, we chose to tie the temperature(s) of the
MEKAL or shock component(s) to be the same in all regions and look for
spatial variations in its (their) intensity (intensities). At a few
locations in the nebula, excess hard X-ray emission was detected and
attributed to HMXBs from circumnuclear star-forming regions. This
contribution was modeled as a cutoff power-law model with cutoff
energy of 10 keV and a fixed $\Gamma$ of 1.1, following Braito et
al. (2004). We assumed no intrinsic absorption to the nebula, i.e.,
the hydrogen column density outside of the nucleus was fixed to the
Galactic value. Finally, lines were identified using the ATOMDB
database (atomdb.org). Line identification was carried out by choosing
the line in the energy range with the expected highest relative
intensity.

In summary, the best-fit model can be described in equation form as:
\begin{eqnarray*}
  {\rm Model} &=& N_{\rm H,~Galactic} \times \{f_{\rm nuclear~NuSTAR} \times N_{\rm
    H,~nucleus} \times ({\rm MYTorus}[N_{\rm H,~AGN}, PL_{\rm AGN}] \\
  & & +~f_{\rm C-thin} \times PL_{\rm AGN} + N_{\rm H,~nuclear~HMXB} \times
  PL^{\rm cutoff}_{\rm nuclear~HMXB}) \\
  & & +~(N_{\rm H,~host~HMXB} \times PL^{\rm cutoff}_{\rm host~HMXB} + {\rm line[1-4] + Nebula})\}, 
\end{eqnarray*}
where $N_{\rm H,~Galactic}$ = 1.26 $\times$ 10$^{20}$ cm$^{-2}$, the
Galactic column density in the direction of Mrk~231 (Dickey \& Lockman
1990), $f_{\rm nuclear~NuSTAR}$ is the fraction ($< 1$) of the NuSTAR
spectrum of Teng et al. (2014), minus the two MEKAL models of the
nuclear diffuse emission, within the given aperture, $N_{\rm
  H,~nucleus}$ is an additional absorbing column in the line of sight
toward the nucleus, not seen in the shallower\footnote{Note that the
  {\em Chandra} data used by Teng et al. (2014) is much shallower than
  what we use here since Teng et al. (2014) only used the {\em
    Chandra} data that were strictly simultaneous with the NuSTAR
  observations -- less than 50 ksec -- to avoid issues associated with
  AGN variability.}  {\em Chandra} spectrum of Teng et al. (2014),
$N_{\rm H,~AGN}$ is the absorbing column of the AGN emission
calculated as part of the MYTorus model and $PL_{\rm AGN}$ is the
direct AGN emission within the MYTorus model that also includes the
scattered fraction and Fe lines, $f_{\rm C-thin}$ is the fraction (=
0.19$^{+0.04}_{-0.03}$) of the ``leaked'' direct AGN emission, $N_{\rm
  H,~nuclear~HMXB}$ $\times$ $PL_{\rm nuclear~HMXB}$ is the highly
obscured emission from HMXBs in the nucleus, $N_{\rm H,~host~HMXB}$
$\times$ $PL_{\rm host~HMXB}$ is the emission from HMXBs outside of
the nucleus (only detected in annulus \#3), line[1 -- 4] are
Gaussian fits to the emission lines, and Nebula = MEKAL$_1$ +
MEKAL$_2$ or vpshock to reproduce the emission from the hot diffuse
gas (vpshock is a constant temperature, plane-parallel shock plasma
model; e.g., Borkowski, Lyerly, \& Reynolds 2001).

\subsubsection{Radial Profile}

The annular spectral extraction regions were defined in the following
fashion: (1) Nucleus –- defined to be $R < $ 1.0 kpc
($\sim$1$\farcs$15). This is a conservatively large radius to make
sure that most of the nuclear emission is included in this region (see
Figure 2). (2) Annulus \#1 (1.0 -- 2.0 kpc) –- this corresponds to the
outflow region from RV11, approximately matching the radial extent of
the Na I~D outflow, but avoiding the bright star-forming arc to the
south. (3) Annulus \#2 (2.0 -- 6.0 kpc) –- this corresponds to the
host galaxy region and includes all of the emission from the southern
star-forming arc. The outer radius was chosen to include only the
portion of the nebula where the PSF wings of the central AGN still
contribute to the hard X-ray flux. This region also corresponds
roughly with the brighter, more relaxed portion of the merger remnant
(e.g., Veilleux et al. 2002, 2006). (4) Annulus \#3 (6.0 -- 16.0 kpc)
-– the outer radius corresponds roughly to the optical edge of the
merger remnant. (5) Annulus \#4 (16.0 -- 40.0 kpc) –- this includes
all of the soft X-ray emission that is outside of the optical remnant
but still largely within the optical tidal complex shown in Koda et
al. (2009).

The extracted X-ray spectra and their best-fit models are shown in
Figure 5 and the derived properties are listed in Table 2.  The shapes
of the spectra show a significant dependence on radial distance,
confirming the radial gradient in the hardness ratio displayed in
Figures 2 -- 3. Our best-fit models translate this hardness ratio
gradient into a radial dependence on the intensity of the central AGN
component, consistent with the expected PSF and our image analysis (\S
3.1), relative to the soft X-ray emission from the extended
nebula. The drop in hardness ratio with increasing radial distance
from the nucleus out to annulus \#2 is due entirely to this effect.
In annulus \#3 ($R$ = 6 - 16 kpc), the flux from the PSF wings of the
central AGN is negligible, yet hard ($>$ 2 keV) X-rays are still
detected. The high-energy portion of this spectrum was fitted using a
cutoff power-law component meant to represent the HMXBs from
circumnuclear star-forming regions. The implied SFR in this annulus is
$\sim$10 $M_\odot$ yr$^{-1}$. This component is not needed in the
outer halo (annulus \#4). The soft X-ray continuum emitted by the
nebula is fit equally well with thermal (MEKAL) and shock models (all
have $\chi^2$/d.o.f. $\la$1.1; see notes to Tables 2a and 2b).
The thermal models require a second MEKAL component with lower
temperature (0.27 keV) to reproduce the softer spectra beyond $R$ = 6
kpc (annuli \#3 and \#4). The absence of a low-temperature MEKAL
component inside of 6 kpc may be due to extra heating in this region
from the circumnuclear star formation, the central quasar
activity, and/or the galactic-scale AGN-driven outflow. We return to
this issue in the next section, where we examine the azimuthal
spectral variations in the inner region of Mrk~231.

The spectra in Figure 5 were also examined for absorption and emission
features. No obvious sign of O~VII 0.7 keV and O~VIII 0.9 keV
absorption edges is seen in any of these spectra. To quantify this
statement, we added an edge at 0.7 keV and another one at 0.9 keV in
the fits to simulate the O~VII and O~VIII features, respectively. The
difference in $\chi^2_\nu$ was found to be insignificant (0.01 for two
degrees of freedom). The tentative detection of the 0.7 keV warm
absorber in the 2000 {\em Chandra} spectrum of the nucleus of Mrk~231
(Figure 9 of Gallagher et al. 2002) is therefore not confirmed here
despite the factor of $\sim$2 smaller uncertainties on the equivalent
width measurements. However, a number of emission lines are present
and provide constraints on the temperature and abundances of O, Mg,
Si, and Fe in the nebula (using the abundances of Wilms, Allen, \&
McCray 2000).  The features around 1.3 and 1.9 keV are identified as
Mg XI 1.352 keV and Si XIII 1.864 keV, respectively.  Both of these
features are present outside of the nucleus: the Si XIII line is
detected in the spectra of annuli \#1, \#2, \#3, and \#4, while Mg XI
is convincingly detected in annuli \#3 and \#4.  In the nuclear
spectrum, there are also two lines near 6.4 and 6.7 keV with
relatively small equivalent widths, consistent with previous data
(e.g., Teng \& Veilleux 2010) and the new analysis of Teng et
al. (2014). These two lines are neutral or weakly ionized Fe~K$\alpha$
and He-like Fe~XXV~K$\alpha$, respectively.  The Fe~XXV 6.7 keV line
may also be present outside of the nucleus (Figure 5). We return to
this point in \S 3.2.5, where we present narrowband line images
derived from our data.

Overall, the results of the fits listed in Table 2 suggest subsolar Fe
abundances and solar or supersolar Si abundances throughout the nebula
(particularly in the inner region; annulus \#1), while O and Mg
abundances fall between these two extremes. These abundances apply
only to the dominant ($\sim$90\%) warmer thermal component of the
halo. The abundances of the cooler component could not be constrained
separately because this component is too faint. In Table 2, the
abundances of the cooler component were fixed to those of the warmer
component.  The absolute values of these abundances should be treated
with caution [e.g., Kim 2012; although our use of a two-temperature
model to fit the halo spectra should reduce the so-called ``Fe bias''
(Buote \& Fabian 1998; Buote 2000)], as implied by the large error
bars on these measurements in Table 2.
However, the relative abundances are considerably more robust. The
relative abundances of the $\alpha$-elements (Si, O, and Mg) with
respect to iron are $\sim$ 2 -- 4 $\times$ solar. Similar supersolar
$\alpha$-element abundances were recently found in the warm thermal
component of the halo of NGC~6240 (Nardini et al. 2013), a galaxy
merger that is caught in the pre-coalescence stage, i.e. at an earlier
stage of evolution than Mrk~231.
We discuss the implications of these results in \S 4.2.

\subsubsection{Neutral Outflow Region}

The spatially resolved Na~ID outflow detected by Rupke et al. (2005c)
and mapped by RV11 and RV13 extends to at least $\sim$2.0 -- 2.5 kpc,
depending on azimuth angle.  There is a region $\sim$1.0 -- 2.5 kpc
due north from the nucleus of Mrk~231 where the Na~ID outflow
velocities are noticeably higher than along other directions (see
right panels in Figure 4 of RV11). RV11 argue that the outflowing
neutral gas in this region may be given an extra kick by the radio jet,
seen on smaller scale along this same direction (e.g., Carilli et
al. 1998; Ulvestad et al. 1999).  With this in mind, we extracted
spectra from four separate quadrants within annulus \#2 ($R = 1.0 -
2.0$ kpc): one coinciding roughly with the possibly jet-influenced
region (PA = $-$45 -- $+$45$^\circ$) and three other regions of the
same radial and azimuthal extent but due east, south, and west (PA =
45 -- 135, 135 -- 225, and 225 -- 315$^\circ$, respectively).  Note
that the radial extent of these regions was limited to 2.0 kpc,
instead of 2.5 kpc, to avoid contaminating emission from the
star-forming arc beyond 2 kpc which would affect the spectra in the
southern comparison region.

The fact that these spectral regions are at the same distance from the
nucleus eliminates any effects that may be due to the radial gradient
of the hardness ratio associated with the wings of the AGN component
discussed in the previous section. The four spectra were binned to at
least 25 counts per bin and fitted simultaneously. We used the same
fitting method here as that used for the annular regions, fitting all
four spectra simultaneously. As in \S 3.2.1, the soft X-ray emission
was modeled using either a single thermal MEKAL component with fixed
$kT$ or a single shock component with fixed $kT$. The strength of this
soft component was left as a free parameter to reflect variations
between the various regions, while the flux from the wings of the PSF
was held fixed. The results are shown in Figure 6 and tabulated in
Table 3 for the various models.

The hardness ratios tabulated in Table 3 indicate that the emission
from the eastern quadrant is slightly softer than that from the other
three regions. RV11 and RV13 have pointed out the presence of a
prominent H~II region $\sim$ 1 kpc due east of the nucleus (see Figure
20 of RV13); this H~II region likely contributes to the excess soft
X-ray emission detected in the eastern spectrum. A fainter H~II region
may also be present in the northern quadrant (see the
low-[N~II]/H$\alpha$ ``blob'' in the middle top panel of Figure 20 of
RV13) and may be contributing to the soft X-ray emission in this
region, although the hardness ratio at that location is not
significantly different from that in the western region.  We see no
evidence (e.g., higher temperatures in Table 3) for jet interaction
with the ISM in the northern region. The spectral fits reveal the
presence of a line at 1.9 keV in the eastern quadrant and perhaps also
in the northern quadrant. This line is identified as Si~XIII 1.864
keV.
The fits require super-solar Si abundances that may reflect
$\alpha$-element enhancement from star formation at these particular
locations. 

Most notably, the fits indicate a clear deficit of soft X-rays in the
western outflow region. This region shows the faintest H$\alpha$
emission and some of the largest Na~ID equivalent widths (Figures 20
and 18d of RV13, respectively).  X-ray absorption by the column of
neutral material at this location ($N_H \sim 10^{22}$ cm$^{-2}$ from
RV13) may explain the weaker soft X-ray flux. Such photoelectric
absorption would harden the soft X-ray spectrum by preferentially
removing the softest X-rays and would thus translate into higher fit
temperatures. To test this possibility, we searched for azimuthal
temperature variations by allowing the temperature of the MEKAL
component to vary for the fits in Table 3. In this case, all quadrants
were assumed to have the same abundances, resulting in Si =
4.7$^{+4.1}_{-2.3}$ solar, Fe = 0.3$^{+0.3}_{-0.2}$ solar, and the
other elements held fixed at solar. The resulting temperatures were
$kT = 0.80^{+0.27}_{-0.19}$ keV, $0.70^{+0.19}_{-0.20}$ keV,
$0.26^{+0.42}_{-0.17}$ keV, and $0.88^{+0.29}_{-0.19}$ keV for the
eastern, southern, western, and northern regions, respectively (the
temperature in the western region is harder to constrain due to the
lower counts in this region). These numbers are all consistent with
each other within the (rather large) uncertainties of these
measurements. The deficit of soft X-rays in the western outflow region
therefore appears to be intrinsic rather than due to photoelectric
absorption by the neutral outflow. We return to this result in \S 4.1.

\subsubsection{Southern Tidal Arc}

Figure 4 shows the extraction windows used to characterize the X-ray
emission from the arc and three comparison regions of the same size
and radial extent (2.0 -- 4.5 kpc) as the arc region but located at
different position angles. The results of the spectral analysis are
shown in Figure 7 and tabulated in Table 4. The soft X-ray emission in
the western arc region is significantly weaker than in the other arc
regions. In contrast, the hard X-ray emission is very nearly constant,
resulting in a higher hardness ratio in the western region. This soft
X-ray flux deficit is analogous to the behavior at smaller radii (\S
3.2.2) and may thus be physically related to the outflow.

The only other significant difference between these spectra is the
detection of emission lines near $\sim$1.8 keV and perhaps also at
$\sim$1.2 keV in the arc region but not in the other regions. These
features are identified as Si XIII 1.864 keV and Ne X 1.211 keV (or Fe
XIX 1.258 keV), respectively. This excess line emission may reflect
$\alpha$-element enhancement due to the starburst in the arc region.

\subsubsection{Northern Tidal Tail}

A polygonal extraction region (Figure 8) was used to extract the 0.5
-- 2 keV spectrum in the region blocked by the northern tidal
tail. The same polygonal footprint was used to extract a comparison
spectrum in a region near the tidal tail, at approximately the same
distance from the center. The best-fit model from annulus \#4 is
scaled to fit the comparison spectrum. Then an absorption component is
added to the best-fit model, with increments of $\Delta N_H$ = 5
$\times$ 10$^{20}$ cm$^{-2}$ at a time, until the total 0.5 -- 2 keV
counts in the simulated spectrum correspond to the total counts from
the region affected by the tidal tail. The hydrogen column density
derived in this manner is 2.5 $\times$ 10$^{21}$ cm$^{-2}$. This
column density is formally a lower limit since foreground emission
would imply a larger column density. The value of this lower limit is
slightly smaller than the column density derived by Iwasawa et
al. (2011) in the tidal tail of Mrk~273, 6 $\times$ 10$^{21}$
cm$^{-2}$, but similar to the typical HI column density of an edge-on
galaxy disk (e.g., Begeman 1989; Barber et al. 1996 have also made
measurements of disk shadowing against the extragalactic X-ray
background). This difference in column densities between the tidal
tails of Mrk~231 and Mrk~273 may simply be an orientation effect: the
tidal tail in Mrk~273 appears thinner, less diffuse than the northern
tidal tidal in Mrk~231, perhaps indicating a more edge-on orientation.

\subsubsection{Narrowband Line Images}

Guided by the detection of emission lines outside of the nucleus
(Figures 5, 6, and 7), we extracted narrowband line images to
investigate the two-dimensional spatial distribution of the
line-emitting gas. The results for Si XIII 1.864 keV, neutral or
slightly ionized Fe K$\alpha$ 6.4 keV, and the sum of He-like Fe XXV
6.7 keV and H-like Fe XXVI 6.9 keV are shown in Figure 9 (Mg XI 1.352
keV was too faint for this exercise). The derived strengths of the
line emission are unreliable in the central 3.0 kpc diameter region
due to Poisson noise from the very strong underlying AGN continuum;
this region is therefore masked in all three panels of Figure 9.

We find clear evidence for Si XIII emission outside of the nucleus,
extending at least $\sim$5 kpc south of the nucleus. Some of the
strongest Si XIII emission coincides with the southern
star-forming arc, confirming the results from our spectral analysis
(\S 3.2.3).

We do not find convincing evidence for Fe K$\alpha$ emission outside of
the nucleus. The small extension to the north-west in the middle panel
of Figure 9 is not statistically significant. A slightly more
significant extension is seen in the Fe XXV 6.7 keV + Fe XXVI 6.9 keV
line emission map (right panel). The spectrum extracted from a
2-arcsec diameter region centered on the brightest part of that
extension is shown in black in Figure 10. The spectrum in red is from
an equal-size region on the opposite side of the nucleus where there
is no obvious Fe XXV + Fe XXVI emission in the line emission map.
There is a difference of 16 counts between these two spectra in the
0.5-7 keV band, but some of this excess in counts could be due to a
slightly higher continuum level.  A line seems to be present in the
black spectrum but not in the red one. Formally the fit shown in
Figure 10 gives $E$(line) = 6.629$^{+0.063}_{-0.044}$ keV, $EW$(line)
= 2.042$^{+1.693}_{-1.743}$, and a narrow width ($\sigma$ = 0 keV
formally). This is thus a tentative detection of Fe XXV 6.7 keV
outside of the nucleus.

The fits to the annular regions discussed in \S 3.2.3 (Figure 5, Table
2) suggested the presence of Fe XXV 6.7 keV in annulus \#3 ($R = 6 -
16$ kpc), where the AGN does not contribute any continuum
emission. This line emission is not immediately obvious in the Fe XXV
+ Fe XXVI emission map. The spectrum extracted from annulus \#3 is
shown in more detail in Figure 11 using a finer binning than in Figure
5. The best fit ($\chi^2_\nu$ = 1.18) shown in this figure constrains
the equivalent width of the Fe XXV 6.7 keV line to be
2.09$^{+1.94}_{-1.40}$ keV. Given the large error bars, the detection
of this line is thus again only tentative.

\section{Discussion}

\subsection{The Quasar-Driven Wind of Mrk~231}

The main objective of the new {\em Chandra} data set was to provide
enough counts to allow us to carry out a detailed spectral imaging
analysis of the galactic-scale outflow region. The results of this
analysis were discussed in detail in \S 3.2.  Everywhere in the
nebula, we find that thermal (MEKAL) models fit the data equally well
as simple shock models. The gas in the outer ($R > 6$ kpc) nebula is
characterized by a dominant ($\sim$90\%) thermal component at $kT \sim
0.7$ keV and a fainter ($\sim$10\%) component at $kT \sim 0.3$ keV. The
cooler component is not present in the inner ($R <$ 6 kpc) region.  No
sign of O~VII 0.7 keV and O~VIII 0.9 keV absorption edges (``warm
absorbers'') is found in any of the spectra, further indicating the
absence of cool gas in the central region of Mrk~231.  This result is
somewhat surprising since warm absorbers are ubiquitous in Seyferts
(e.g., Crenshaw et al. 2003) and QSOs (e.g., Piconcelli et al. 2005;
Teng \& Veilleux 2010).

A more detailed study of the outflow region, where we sliced the
annular region between $R$ = 1 and 2 kpc into four equal-size
quadrants (N, S, E, and W), reveals significant variations in hardness
ratios and intensities. The spectrum in the eastern outflow region is
found to be significantly softer and shows a larger Si/Fe abundance
ratio than the other outflow regions. This is likely due to the
presence of an HII region visible in the optical data (RV13). A
fainter HII region may also be affecting the spectrum in the northern
quadrant. A deficit of soft X-ray emission is present in the western
quadrant, coincident with fainter H$\alpha$ emission and some of the
largest neutral-gas Na~ID absorption columns. This is the strongest
evidence in our data that the hot X-ray emitting gas ``knows'' about
the massive neutral outflow in this object. There is also tentative
evidence for Fe XXV 6.7 keV line emission extending up to $\sim$3 kpc
north-west of the nucleus, but this is only a 2-sigma detection so it
is not very significant (Figures 9 -- 10).

It is perhaps surprising that our data do not show any obvious
temperature enhancements in the neutral outflow region given that the
velocities of the outflowing neutral gas are typically $\sim$1000 km
s$^{-1}$. One would naively expect this neutral outflow to interact
with the ambient ISM and produce strong ionizing shocks. For
non-radiative, strong shocks in a fully ionized monoatomic gas, the
post-shock temperature $T_{sh} = 3 \mu v^2_{sh}/16 k$, where $\mu$ is
the mean mass per particle and $k$ is the Boltzmann constant (McKee \&
Hollenback 1980). For shock velocities of $\sim$1000 km s$^{-1}$, we
expect hot gas with $T \sim$ 1.6 $\times$ 10$^7$ K or $kT \sim$ 1.40
keV. The tentative detection of extended Fe XXV 6.7 keV emission would
imply shock velocities of $\ga$2000 km s$^{-1}$, if produced by a
collisionally ionized plasma with a temperature $T \sim 7 \times 10^7$
K as in NGC~6240 (Wang et al. 2014). These high velocities are not
seen in the neutral gas.  Our fits do not formally rule out the
possibility that HMXBs at this location may be responsible for this
extended Fe XXV emission.

These simple back-of-the-envelope arguments emphasize the fact that
one has to be very cautious when using the properties of the neutral
outflow to predict those of the hot X-ray emitting material. This may
not be too surprising given the huge disparity in temperatures
($<$1000 K versus $\sim$10$^7$ K) and densities ($>$1 cm$^{-3}$ versus
$\sim$10$^{-2}$ cm$^{-3}$; Table 3) between these two gas phases. It
would be more prudent to compare the properties of the X-ray emitting
gas with those of the warm ionized gas traced by H$\alpha$, which is
often found to be spatially correlated with the soft X-ray emission
(e.g., Cecil, Bland-Hawthorn, \& Veilleux 2002; Strickland et
al. 2004; Bolatto et al. 2013).  Unfortunately, contaminating
H$\alpha$ emission from HII regions complicates the picture in Mrk~231
and the H$\alpha$ outflow is detected convincingly only within $\sim$1
kpc (RV13).  Indeed, all of the evidence (RV11, RV13, Fischer et
al. 2010; Feruglio et al. 2010; Gonzalez-Alfonso et al. 2014) suggests
that the outflow of Mrk 231 is heavily mass-loaded with neutral and
molecular material but relatively little ionized gas; our new {\em
  Chandra} data add support to this idea. The absence of hot-shocked
ionized gas in the wind region naively suggests that the outflow is
momentun-driven rather than energy-driven. However, this is not a
strong conclusion since we cannot strictly rule out the presence of a
very hot ($>>$10$^7$ K) and tenuous wind fluid gas (present in the
energy-driven wind of M82; Griffiths et al. 2000; Stevens, Read, \&
Bravo-Guerrero 2003).

The absence of the cooler thermal component within $\sim$6 kpc
indicates either more efficient cooling on large scale or additional
heating on small scale. Adiabatic expansion of a free wind can
naturally explain negative temperature gradients but, as we just
mentioned, there is very little evidence that any of the soft X-ray
emitting gas detected within $\sim$6 kpc is directly associated with
the wind event. Moreover, there is no evidence at present that the
galactic wind in Mrk~231 extends beyond $\sim$3 kpc (RV11, RV13), let
alone $>$6 kpc (note however that this is purely a limitation of the
current optical data: Na ID absorption is not detected beyond $\sim$3
kpc because the galaxy continuum is too faint). Even if the wind did
extend beyond 6 kpc, it is not clear how the wind scenario could
explain the presence of the (dominant) warmer component on the largest
scale (out to $\sim$25 kpc from the nucleus). Finally, the drop in
X-ray surface brightness in Figure 2 is also less steep than the $\sim
R^{-3}$ profile expected in the case of a freely expanding wind (where
the gas and electron density profiles $n_g$ and $n_e$ go as $\sim
r^{-2}$, from mass conservation, and the X-ray surface brightness
$\Sigma_X \sim n_e^2 dV/dS$, where $dV$ stands for the volume element
and $dS$ is the surface element).  We therefore favor the scenario
where the absence of the cooler component within 6 kpc is due to extra
heating.

Known sources of heating in this region include the galactic-scale
wind, the circumnuclear starburst, and the quasar itself. We discuss
each in turn.  The total kinetic energy of the AGN-driven
galactic-scale outflow in Mrk 231 is substantial ($\sim$ 6 $\times$
10$^{57}$ ergs) and dominated by the contributions from the neutral
and molecular components (RV11, RV13, Feruglio et al. 2010, Cicone et
al. 2012, and Gonzalez-Alfonso et al. 2014).  For comparison, the {\em
  total} thermal energy of the X-ray emitting gas in the inner ($R <$
6 kpc) region of Mrk~231 is $3 \eta n_e V k T \approx 10^{57}$ ergs,
where we used a filling factor $\eta \sim 1$ (volume-filling), an
electron density $n_e$ $\sim$ 2 $\times$ 10$^{-2}$ cm$^{-3}$ (Table 3;
$\propto \eta^{-1/2}$), a spherical volume $V \sim 2.4 \times 10^{67}$
cm$^3$, and a temperature of 8 $\times 10^6$ K (0.7 keV). The absence
of cooler ($\sim$0.3 keV) X-ray emitting gas in the inner region of
Mrk~231 may therefore be due to the influence of the AGN-driven
outflow on the ambient ISM, even if only a small fraction of this
kinetic energy is thermalized via shocks (e.g., Cox et al. 2006).
However, the lack of a clear signature of ionizing shocks in the
neutral outflow region and the current size limits on this outflow
($\ga$3 kpc) weaken this wind scenario.

The star formation rate of Mrk~231 is $\sim$140 $M_\odot$ yr$^{-1}$
(Veilleux et al. 2009). This starburst therefore injects mechanical
energy in the surrounding medium at a rate $\sim$1 $\times$ 10$^{44}$
erg s$^{-1}$ (e.g., Veilleux et al. 2005). The age of this starburst
is not well constrained. Assuming a conservatively small value of $5
\times 10^6$ yrs for the starburst age, the starburst can contribute
up to 10$^{57}$ ergs if the thermalization efficiency is $>$10\%. This
is more than enough to explain the lack of a cool component in the
inner 6 kpc. However, there is no evidence that the starburst extends
much beyond $\sim$3 kpc, except in the southern tidal arc where the
starburst may be blowing its own low-velocity wind (RV11, RV13).

The bolometric luminosity of the quasar in Mrk~231 is $\sim$3 $\times$
10$^{12}$ $L_\odot$ (Veilleux et al. 2009), but most of this energy is
emitted in the infrared and cannot contribute to heating the hot X-ray
emitting gas in the inner 6 kpc of the nucleus. The recent {\em
  NuSTAR} data of Teng et al. (2014) indicate that Mrk~231 is
underluminous in the X-rays with an intrinsic absorption-corrected 0.5
-- 30 keV luminosity of 1.0 $\times$ 10$^{43}$ erg s$^{-1}$. Assuming
a quasar lifetime of 10$^7 - 10^8$ yrs (see, e.g., review by Martini
2004), only $\sim$2 -- 20\% of the 0.5 -- 30 keV total energy output
from the quasar in Mrk 231 needs to be absorbed within the inner 6 kpc
of the quasar to contribute $\sim$ 10$^{57}$ ergs to the heating of
the inner region. The analysis of the {\em NuSTAR} data indicates that
a patchy and Compton-thin ($N_H \sim 1.12^{+0.29}_{-0.24} \times
10^{23}$ cm$^{-2}$) column absorbs (1 - C-thin) = $\sim$80\% of the
intrinsic 0.5 -- 30 keV luminosity of the quasar in Mrk~231. Assuming
that our line of sight to the quasar is typical of other directions,
this energy is amply sufficient to explain the lack of a cool
component in the inner 6 kpc of Mrk 231. However, this energy is not
deposited at the right place. The absorbing column measured by {\em
  NuSTAR} is spatially unresolved ($<$ 1 kpc) so the energy absorbed
by this material is deposited on a scale considerably smaller than 6
kpc. Perhaps some fraction of the quasar energy output that makes it
out of the nucleus is absorbed by material on $>$kpc scale, including
possibly the neutral atomic and molecular gas entrained in the
wide-angle outflow (large-scale ionization is seen in the
circumgalactic nebula of MR2251$-$178 for instance; Kreimeyer \&
Veilleux 2013). Another possibility is that the energy deposited on
sub-kpc scale is redistributed on larger scales via slower dynamical
processes, e.g., buoyant bubbles of hot gas mixing with the cooler
X-ray material, as seen in the intracluster medium of a growing number
of galaxy clusters (e.g., Hlavacek-Larrondo et al. 2012).

Our current data set does not allow us to discriminate between heating
by the galactic wind, the circunuclear starburst, and the quasar.
Most likely, all three contribute at some level to the absence of the
cool thermal component within $\sim$6 kpc.

\subsection{Origin of the X-ray Halo}

The spectacular X-ray halo in Mrk~231 has many properties that are
similar to those of the halo in NGC~6240 (Nardini et al. 2013), a
galaxy merger at an earlier stage of evolution than Mrk~231. We argue
below that they likely have similar origins.

The total luminosity of the nebula around Mrk 231 is $\sim$2 $\times$
10$^{41}$ erg s$^{-1}$ (annuli \#1 -- \#4 in Table 2, excluding the
contribution from the wings of the AGN PSF and host HMXBs), similar to
that of the halo in NGC~6240 and, as pointed by Nardini et al. (2003),
comparable to that of small groups of galaxies (Mulchaey 2000) and
giant ellipticals (Canizares et al. 1987; Mathews \& Brighenti 2003).
The full halo spectrum of NGC~6240 calls for two thermal (MEKAL)
components with temperatures $kT \sim$ 0.8 and 0.25 keV, while our
best-fit model for the halo in Mrk~231 involves two thermal components
with $kT$ $\sim$0.7 keV and $\sim$0.3 keV with the latter component
contributing only $\sim$10\% of the luminosity. Given the smaller
dimensions of the nebula around Mrk~231 ($\sim$65 $\times$ 50 kpc)
relative to that of NGC~6240 ($\sim$110 $\times$ 80 kpc), the
sound-crossing time, $D/c_s = D (5 kT/3 \mu m_p)^{-1/2}$, a rough
measure of the dynamical age of the halo, is proportionally smaller in
Mrk~231: $\sim$100 Myr versus $\sim$200 Myr. The thermal energy
content of the halo in Mrk 231, $E_{th} \sim 2 \times 10^{58}$ erg, is
also slightly smaller than that of the halo in NGC~6240 ($\sim$5
$\times$ 10$^{58}$ erg), but still considerable if we compare it to
the amount of kinetic energy deposited during the merger of two
identical progenitors, of order $M_g v_c^2/8$, where $M_g$ is the mass
of the X-ray emitting gas and $v_c$ is the relative speed during the
collision (Nardini et al. 2013). In the case of Mrk~231, $M_g = \eta
n_e V m_p \sim 7 \times 10^{9}$ $M_\odot$ so $v_c \sim 1400$ km
s$^{-1}$ is needed, which is subtantially faster than typical head-on
collisions in non-cluster environments (e.g., Taffy Galaxies; Braine
et al. 2003). A comparison of the thermal energy content of the halo
in Mrk 231 with its X-ray luminosity implies a cooling timescale
$\sim$10$^{17}$ sec or $\sim$3 Gyr.

As discussed in \S 4.1, the present-day starburst, AGN, and galactic
wind in the core of Mrk~231 no doubt have injected some energy in the
X-ray nebula and may be responsible for the absence of cool gas in the
inner ($<$ 6 kpc) region of the nebula. But, they are unlikely to have
supplied the entire thermal energy content of the X-ray halo given the
energetics of these processes (see \S 4.1). The large size of the
nebula also puts severe constraints on the duration of these events:
e.g., $\ga$100 Myr for an average velocity of $\la$500 km s$^{-1}$,
which seems at odd with the starburst age ($<$10$^7$ yrs) and the
dynamical age of the present-day galactic wind (a few $\times$ 10$^6$
yrs).

Independent constraints on the processes at work come from our
abundance analysis of the halo (annuli \#3 and \#4 in \S 3.2.1; Tables
2a and 2b). We measure $\alpha$-element (particularly Si, but also O
and Mg) abundances relative to iron that are $\sim$2 -- 4 $\times$
solar throughout the nebula without significant radial gradient.  A
similar abundance pattern was found in the halo of NGC~6240 (Nardini
et al. 2013). Iron is mainly produced from type Ia supernovae (SNe;
i.e., exploded white dwarfs in close binary systems) on a timescale of
$\sim$0.1-1 Gyr, while the $\alpha$-elements come primarily from type
II SNe (i.e., core-collapsed massive stars) on timescales of a few
tens of Myr.  Synthesis models for Type II SNe (e.g, Heger \& Woosley
2010; Nomoto et al. 2006, 2013) predict Si/Fe ratios of up to
$\sim$3-5 solar, while ratios of $\sim$0.5 solar are expected from
Type Ia SNe (e.g., Nagataki \& Sato 1998; Seitenzahl et al. 2013).
The supersolar Si/Fe ratios in the halos of Mrk~231 and NGC~6240
therefore suggest uniform enrichment by type II SNe out to the largest
scales ($\sim$65 $\times$ 50 kpc in the case of Mrk~231 and 110
$\times$ 80 kpc for NGC 6240). Supersolar Si/Fe ratios have been found
in the past [e.g., face-on spiral galaxies (Schlegel et al. 2003;
Soria \& Wu 2003), galaxy mergers like the Antennae (Baldi et
al. 2006a, 2006b), and the central regions of young elliptical
galaxies with sites of recent (a few tens of Myr) merger-induced star
formation (Kim et al. 2012)], but in all of these cases, the
supersolar Si/Fe ratios are measured on kpc scale, not several tens of
kpc as in Mrk~231 and NGC~6240.

In Mrk~231, we see evidence for $\alpha$-element enhancements near and
around the circumnuclear starburst (\S 3.2.2) but the on-going (age
$<$10$^7$ yrs) episode of active star formation ($SFR \sim$ 140
$M_\odot$ yr$^{-1}$) does not seem capable of explaining the
large-scale enhancements seen in the halo. The estimated mass in
silicon in the halo is $\sim$6 $\times$ 10$^6$ $M_\odot$, assuming a
solar silicon abundance (Table 2a).
The maximum silicon yield of type II SNe is $\sim$0.1 -- 0.3 $M_\odot$
for a massive-star progenitor with $Z \le 0.02$ (e.g. yields tables in
Nomoto et al. 2013) and thus implies the need for $\sim$3 $\times$
10$^7$ type II SNe or a sustained star formation rate of $\sim$140
$M_\odot$ yr$^{-1}$ over $\ge$10$^7$ yrs. Moreover, the metals
produced in the circumnuclear starburst need to be redistributed over
the entire $\sim$55 $\times$ 60 kpc nebula to be consistent with the
observations. In some cases, there is direct evidence that galactic
winds help carry the $\alpha$-element enhanced material produced by
starburst into the halos of galaxies (e.g., M82: Tsuru et al. 2007;
Ranalli et al. 2008; Konami et al. 2011 and references therein), but
the present-day galactic wind in Mrk~231 cannot be responsible for the
supersolar Si/Fe ratios throughout the halo of Mrk 231 unless the wind
actually extends $\sim$1 order of magnitude further than currently
measured ($\ga$3 kpc, RV11, RV13). Another, more likely, scenario is
that the $\alpha$-elements produced near the center have been
redistributed on larger scale by previous outflow events. 

In NGC~6240, the warmer ($kT_1 \sim$ 0.8 keV) component of the halo is
distinctly metal-richer ($Z_\alpha \sim$ 0.5 solar) than the cooler
($kT_2 \sim 0.25$ keV) component (fixed at $Z = 0.1$ solar). Nardini
et al. (2013) associate the first component with chemically-evolved,
starburst-processed gas and the second component with
gravitationally-bound, pre-existing halo material. The cooler halo
component in Mrk~231 is $\sim$10$\times$ weaker than the warmer
component so, unfortunately, our {\em Chandra} data do not have
sufficient counts to constrain the metal abundance in this fainter
component, hence its origin\footnote{For instance, we determined that
  models with a low metal abundance of $Z = 0.1$ solar in the cooler
  component fit the data equally well as models with $Z = 1$
  solar.}. However, regardless of the exact origin of the cooler halo
component in Mrk~231, we mentioned in the previous paragraph that the
sheer amount of metals in the warmer component and its widespread
supersolar Si/Fe ratios point to enhanced star formation over
timescales comparable to the dynamical time ($\sim$100 Myr) to both
produce these metals and redistribute them across the halo via outflow
events. The merger itself may also help in redistributing the
metals. Mrk~231 is in the final throes of a major merger of two
gas-rich disk galaxies with masses similar those of the Milky Way and
Andromeda (Cox \& Loeb 2008). The colliding gas in the parent disk
galaxies of the merger is shock-heated to X-ray-emitting temperatures
and eventually mixed with the pre-existing halo material to contribute
to the observed X-ray halo (Jog \& Solomon 1992; Cox et al. 2006).
The lack of systematic Si/Fe ratio variations across the halo of
Mrk~231 suggests that the collision and outflow events were efficient
at erasing abundance gradients.

Some fraction of the X-ray gas present in the halo of Mrk~231 will
likely be retained by the merger remnant and become the X-ray-emitting
halo of the resulting young elliptical galaxy.  While solar Si/Fe
ratios are typically seen in the X-ray halos of present-day large
elliptical galaxies (e.g,. Humphrey \& Buote 2006; Loewenstein \&
Davis 2010, 2012; Konami et al. 2014), supersolar Si/Fe ratios are
often measured in the stellar components of $>$L$^*$ ellipticals
(e.g., Worthey 1998; Graves et al. 2010; Conroy et al. 2014). The
stellar abundances are generally explained by invoking a short
timescale of star formation (efficient quenching before the onset of
type Ia SNe) or variations in the IMF.  Given the mass outflow rate
and gas content of Mrk~231, the implied gas depletion time scale is
only 10 -- 20 Myr (Sturm et al. 2011; Gonzalez-Alfonso et
al. 2014). The outflow in Mrk~231 could thus quench star formation on
this short time scale, if the ejected gas does not return to the
center to form stars. This type of life-changing outflow event appears
to be common among local ULIRGs (Veilleux et al. 2013; Spoon et
al. 2013; Cicone et al. 2014). Assuming Mrk 231-like outflow events
are also common in major mergers in the early universe, they might
help explain the enhancement of $\alpha$-element, relative to iron,
observed in the stellar population of local large elliptical galaxies.

\section{Conclusions}

We have combined a deep 400-ksec ACIS-S observation of Mrk~231 with
120-ksec archival data acquired with the same instrument and setup to
allow us to carry out the first spatially resolved spectral analysis
of a hot X-ray emitting circumgalactic nebula around a quasar. Mrk~231
is particularly well suited for this study since it is the nearest
quasar and is well known to host a powerful galactic-scale
outflow. The main results from our study are the followings:

\begin{itemize}

\item The morphology of the $\sim$65 $\times$ 50 kpc X-ray nebula does
  not resemble that of the tidal complex of Mrk~231 seen at optical
  wavelengths.  The only spatial match is the small tidal arc located
  $\sim$3.5 kpc south of the nucleus, where excess soft X-ray
  continuum emission and Si XIII 1.8 keV line emission are detected,
  consistent with star formation and its associated $\alpha$-element
  enhancement, respectively. We also detect a deficit in the soft
  X-ray flux map at the position of the northern tidal tail,
  suggesting that this structure casts an X-ray shadow due to a
  hydrogen column density of at least 2.5 $\times$ 10$^{21}$
  cm$^{-2}$.

\item The soft X-ray spectrum of the nebula beyond 6 kpc is best
  described as the sum of two thermal components of temperatures
  $\sim$3 and $\sim$8 million K with spatially uniform super-solar
  $\alpha$ element abundances, relative to iron. A similar result was
  recently found in the X-ray halo of the pre-merger NGC~6240.
  Enhanced star formation activity over an extended period of time
  ($\sim$10$^8$ yrs) is needed to produce the vast amount of $\alpha$
  elements detected in the nebula of Mrk~231. Multiple outflow events,
  such as the on-going quasar-driven galactic wind, may help carry the
  $\alpha$ elements produced in the circumnuclear region out to the
  largest scales. Such wind-driven metal transport is directly seen to
  take place in the nearby starburst M~82, albeit on a considerably
  smaller scale. The stirring of the gas associated with the merger
  itself may also redistribute the metals across the nebula and help
  erase remaining abundance gradients.

\item The hard X-ray continuum emission in the inner ($\le$6 kpc)
  nebula is consistent with being due entirely to the bright central
  quasar and the wings of the {\em Chandra} point spread function.
  The $\sim$3 million K thermal component detected in the halo is not
  present within 6 kpc of the nucleus. No sign of O~VII 0.7 keV and
  O~VIII 0.9 keV absorption edges (``warm absorbers'') is found in any
  of the spectra, further indicating a lack of cool X-ray detectable
  gas in the central region of Mrk~231.  Energetically, heating from
  the circumnuclear starburst, the central quasar activity, or the
  wide-angle quasar-driven outflow is each capable of explaining this
  lack of cool gas. The strongest evidence in our data that the hot
  X-ray emitting gas ``knows'' about the massive neutral/molecular
  outflow in this object is a deficit of soft X-ray emission in the
  western quadrant extending 1 -- 2 kpc (perhaps as far as 2 -- 4.5
  kpc) from the nucleus. This region coincides with fainter H$\alpha$
  emission and some of the largest columns of outflowing neutral gas
  probed by observations of the Na I optical doublet. Shocks created
  by the interaction of the wind with the ambient ISM may heat the gas
  to high temperatures at this location.  Indeed, there is tantalizing
  (2-$\sigma$) evidence for Fe XXV 6.7 keV line-emitting gas extending
  up to $\sim$3 kpc north-west of the nucleus. If produced by a
  collisionally ionized plasma with a temperature $T \sim 7 \times
  10^7$ K, as in the case of NGC~6240, this would imply shock
  velocities $\ga$2000 km s$^{-1}$. HMXBs may also be responsible for
  this extended line emission.
\end{itemize}

\acknowledgements We thank the anonymous referee for thoughtful and
constructive comments that improved this paper. Support for this work
was provided by NASA through {\em Chandra} contract GO2-13129X (S.V.)
and the NASA Postdoctoral Program (NPP) Fellowship (S.H.T., S.V.).
S.V. acknowledges support from the Alexander von Humboldt Foundation
for a ``renewed visit'' to Germany, and thanks the host institution,
MPE Garching, where a portion of this paper was written. He is also
grateful to R. Mushotzky for discussions of the interpretation of the
elemental abundances. This work has made use of NASA's Astrophysics
Data System Abstract Service and the NASA/IPAC Extragalactic Database
(NED), which is operated by the Jet Propulsion Laboratory, California
Institute of Technology, under contract with the National Aeronautics
and Space Administration.

\clearpage

\begin{deluxetable}{ccc}
%\rotate
\tablecolumns{3}
%\tabletypesize{\tiny}
\setlength{\tabcolsep}{0.01in}
\tablecaption{Two-Component $\beta$-Model Fits to the X-ray Surface Brightness Radial Profiles}
\tablewidth{0pt}
\tablehead{\colhead{Model} & \colhead{Inner} & \colhead{Outer}\\
\colhead{Parameter~~} &\colhead{Component~~}&\colhead{Component~~}
}
\startdata
\cutinhead{0.5--2~keV}
R$_0$&0.62&3.31\\
$\beta$&1.00&0.54\\
Amplitude&1421.61&7.88\\
$\chi^2_\nu$&\multicolumn{2}{c}{1.94}\\
\cutinhead{2--7~keV}
R$_0$&0.55&0.03\\
$\beta$&1.00&0.31\\
Amplitude&2481.61&14.13\\
$\chi^2_\nu$&\multicolumn{2}{c}{1.89}\\
\enddata
\label{tab:beta}
\end{deluxetable}

\tablenum{2}
%\begin{turnpage}
\begin{deluxetable}{cccccc}
%\rotate
\tablecolumns{5}
\tabletypesize{\tiny}
\setlength{\tabcolsep}{0.01in}
\tablecaption{a) Simultaneous Model Fit to the Annular Regions (MEKAL models)}
\tablewidth{0pt}
\tablehead{\colhead{Model} & \colhead{Nucleus} & \colhead{Annulus 1} &\colhead{Annulus 2}& \colhead{Annulus 3} &\colhead{Annulus 4}\\
\colhead{Parameter\tablenotemark{a}} &&&&&
}
\startdata
Radial Distance (kpc)&$<$1.0&1.0--2.0&2.0--6.0&6.0--16.0&16.0--40.0\\
Hardness Ratio & 0.260$\pm$0.008&$-0.107\pm0.028$&$-0.648\pm0.025$&$-0.636\pm0.020$&$-1.000^{+0.024}_{-0.000}$\\
Ratio to Nuclear NuSTAR Model &0.8134$^{+0.0132}_{-0.0137}$&
0.0409$^{+0.0029}_{-0.0029}$&0.0257$^{+0.0026}_{-0.0026}$&0 (f)& 0 (f)\\
N$_{\rm H, nucleus}$
(10$^{22}$~cm$^{-2}$)&0.08$^{+0.02}_{-0.02}$&0.08 (t)
& 0.08 (t) &\nodata&\nodata\\
N$_{\rm H, AGN}$
(10$^{24}$~cm$^{-2}$)&0.112 (f) & 0.112 (t) & 0.112
(t) &\nodata&\nodata\\
C-thin
Fraction&0.194 (f) & 0.194 (t) & 0.194 (t) &\nodata&\nodata\\
N$_{\rm H, host~HMXB}$
(10$^{22}$~cm$^{-2}$)&\nodata&\nodata&\nodata& 7.92 (t) &\nodata\\
kT$_1$
(keV)&0.67$_{-0.03}^{+0.03}$& 0.67 (t) & 0.67 (t) & 0.67 (t) & 0.67 (t)\\
n$_{\rm e, 1}$ (cm$^{-3}$) & 0.00 &$4.1 \times 10^{-2}$&$1.5\times10^{-2}$&$4.0\times10^{-3}$&$9.7\times10^{-4}$\\
kT$_2$
(keV)&0.27$^{+0.08}_{-0.27}$& 0.27 (t) & 0.27 (t) & 0.27 (t) & 0.27 (t)\\
n$_{\rm e, 2}$ (cm$^{-3}$) & 0.00 &0.00&0.00&$2.7\times10^{-3}$&$9.2\times10^{-4}$\\
E$_{\rm line, 1}$
(keV)&0.879$_{-0.015}^{+0.024}$&\nodata&\nodata&\nodata&\nodata\\
EW$_{\rm line, 1}$ (keV)&0.038$^{+0.011}_{-0.013}$&\nodata&\nodata&\nodata&\nodata\\
Line ID& Fe XVIII&\nodata&\nodata&\nodata&\nodata\\
E$_{\rm line, 2}$
(keV)&1.003$_{-0.010}^{+0.019}$&\nodata&\nodata&\nodata&\nodata\\
EW$_{\rm line, 2}$
(keV)&0.053$^{+0.013}_{-0.013}$&\nodata&\nodata&\nodata&\nodata\\
Line ID& Fe XXI&\nodata&\nodata&\nodata&\nodata\\
E$_{\rm line, 3}$(keV)&2.085$_{-0.090}^{+0.085}$&\nodata&\nodata&\nodata&1.854$^{+0.041}_{-0.050}$\\
$\sigma_{\rm line 3}$
(keV)&0.242$^{+0.074}_{-0.063}$&\nodata&\nodata&\nodata&0 (f)\\
EW$_{\rm line, 3}$ (keV)&0.133$^{+0.045}_{-0.045}$&\nodata&\nodata&\nodata&0.368$^{+0.358}_{-0.272}$\\
Line ID&Si XIV&\nodata&\nodata&\nodata&Si XIII\\
E$_{\rm line, 4}$
(keV)&6.687$^{+0.034}_{-0.045}$&\nodata&\nodata&6.687 (t)&\nodata\\
EW$_{\rm line, 4}$ (keV)&0.121$^{+0.040}_{-0.034}$&\nodata&\nodata&1.431$^{+2.211}_{-1.355}$&\nodata\\
Line ID&Fe XXV&\nodata&\nodata&Fe XXV&\nodata\\
O/O$_\odot$\tablenotemark{b}&\nodata&0.9$^{+1.8}_{-0.7}$&1.1$^{+0.6}_{-0.4}$&0.5$^{+0.7}_{-0.2}$&0.6$^{+2.3}_{-0.3}$\\
Mg/Mg$_\odot$\tablenotemark{b}&\nodata&0.5$^{+1.8}_{-0.5}$&0.9$^{+0.5}_{-0.4}$&1.2$^{+0.5}_{-0.4}$&0.9$^{+0.7}_{-0.5}$\\
Si/Si$_\odot$\tablenotemark{b}&\nodata&4.5$^{+5.5}_{-2.2}$&1.3$^{+0.7}_{-0.6}$ & 1.5$^{+0.7}_{-0.6}$&1 (f)\\
Fe/Fe$_\odot$\tablenotemark{b}&\nodata&0.3$^{+0.5}_{-0.2}$& 0.4$^{+0.2}_{-0.1}$ & 0.3$^{+0.2}_{-0.1}$&0.3$^{+0.4}_{-0.1}$\\
f$_{\rm 0.5-2~keV}$ (10$^{-14}$~erg~s$^{-1}$~cm$^{-2}$)\tablenotemark{c}&5.26$_{-0.11}^{+0.10}$&0.56$_{-0.10}^{+0.03}$&1.38$_{-0.10}^{+0.04}$&2.03$_{-0.10}^{+0.07}$&2.09$^{+0.10}_{-0.41}$\\
f$_{\rm 2-10~keV}$ (10$^{-14}$~erg~s$^{-1}$~cm$^{-2}$)\tablenotemark{c}&74.47$_{-1.20}^{+1.26}$&3.69$_{-0.25}^{+0.25}$&2.38$_{-0.23}^{+0.24}$&1.98$_{-0.41}^{+0.37}$&0.08$^{+0.02}_{-0.02}$\\
L$_{\rm MEKAL, 1}$
(10$^{40}$~erg~s$^{-1}$)\tablenotemark{d}&0.00&1.38&5.20&6.52&6.00\\
L$_{\rm MEKAL, 2}$
(10$^{40}$~erg~s$^{-1}$)\tablenotemark{d}&0.00&0.00&0.00&2.10&2.47\\
L$^{\rm HMXB}_{\rm host}$ (10$^{40}$~erg~s$^{-1}$)&\nodata&\nodata&\nodata&10.69&\nodata\\
L$_{\rm AGN, 0.5-2
  keV}$(10$^{41}$~erg~s$^{-1}$)\tablenotemark{d}&9.92&0.50&0.32&0.00&0.00\\
L$_{\rm AGN, 2-10
  keV}$(10$^{41}$~erg~s$^{-1}$)\tablenotemark{d}&28.56&1.44&0.90&0.00&0.00\\
\enddata

\tablenotetext{a}{Best-fit model: $N_{\rm H,~Galactic} \times \{f_{\rm
    nuclear~NuSTAR} \times N_{\rm H,~nucleus} \times ({\rm
    MYTorus}[N_{\rm H,~AGN}, PL_{\rm AGN}] + f_{\rm C-thin} \times
  PL_{\rm AGN} + N_{\rm H,~nuclear~HMXB} \times PL^{\rm cutoff}_{\rm
    nuclear~HMXB}) + (N_{\rm H,~host~HMXB} \times PL^{\rm cutoff}_{\rm
    host~HMXB} + {\rm line[1-4]} + {\rm MEKAL}_1 + {\rm MEKAL}_2)\}$,
    where $N_{\rm H,~Galalactic}$ = 1.26 $\times$ 10$^{20}$ cm$^{-2}$,
    the Galactic column density in the direction of Mrk~231 (Dickey \&
    Lockman 1990), $f_{\rm nuclear~NuSTAR}$ is the fraction ($< 1$) of the
    NuSTAR spectrum of Teng et al. (2014), minus the two MEKAL models
    of the nuclear diffuse emission, within the given aperture,
    $N_{\rm H,~nucleus}$ is an additional absorbing column in the line
    of sight toward the nucleus, not seen in the shallow {\em Chandra}
    spectrum of Teng et al. (2014), $N_{\rm H,~AGN}$ is the absorbing
    column of the AGN emission calculated as part of the MYTorus model
    and $PL_{\rm AGN}$ is the direct AGN emission within the MYTorus
    model that also includes the scattered fraction and Fe lines,
    $f_{\rm C-thin}$ is the fraction (= 0.19) of the ``leaked'' direct AGN
    emission, $N_{\rm H,~nuclear~HMXB}$ $\times$ $PL^{\rm cutoff}_{\rm
      nuclear~HMXB}$ is the highly obscured emission from HMXBs in the
    nucleus, $N_{\rm H,~host~HMXB}$ $\times$ $PL^{\rm cutoff}_{\rm
      host~HMXB}$ is the emission from HMXBs outside of the nucleus
    (only detected in annulus \#3), line[1 -- 4] are Gaussian fits to the
    emission lines, and MEKAL$_1$ and MEKAL$_2$ are two MEKAL models
    of the emission from the hot diffuse gas (see text for more
    detail). This best-fit model has $\chi^2$ of 468.7 for 426 degrees
    of freedom ($\chi^2_\nu$=1.10). (t) indicates that the value of
    the parameter is tied to be the same in all regions.  (f)
    indicates that the value of the parameter is held fixed.  }

\tablenotetext{b}{Atomic abundance relative to Solar value.  All other
  elemental abundances fixed at Solar.}  

\tablenotetext{c}{Observed flux.}

\tablenotetext{d}{Absorption corrected luminosity.}

\label{tab:fits}
\end{deluxetable}
%\end{turnpage}

\tablenum{2}
\begin{deluxetable}{ccccc}
%\rotate
\tablecolumns{5}
\tabletypesize{\tiny}
\setlength{\tabcolsep}{0.01in}
\tablecaption{b) Simultaneous Model Fit to the Annular Regions (shock models)}
\tablewidth{0pt}
\tablehead{\colhead{Model} & \colhead{Annulus 1} & \colhead{Annulus 2}
  &\colhead{Annulus 3}& \colhead{Annulus 4}\\
\colhead{Parameter\tablenotemark{a}} &&&&
}
\startdata
Radial Distance (kpc)&1.0--2.0&2.0--6.0&6.0--16.0&16.0--40.0\\
Hardness Ratio & $-0.107\pm0.028$&$-0.648\pm0.025$&$-0.636\pm0.020$&$-1.000^{+0.024}_{-0.000}$\\
Ratio to Nuclear NuSTAR Model
&0.0408$^{+0.0028}_{-0.0028}$&0.0268$^{+0.0025}_{-0.0025}$&0 (f)&0 (f)\\
N$_{\rm H, AGN}$
(10$^{24}$~cm$^{-2}$)&0.112 (f)& 0.112 (t) &\nodata&\nodata\\
C-thin
Fraction&0.194 (f) & 0.194 (t) &\nodata&\nodata\\
N$_{\rm H, host~HMXB}$ (10$^{22}$~cm$^{-2}$) &\nodata&\nodata& 7.55$^{+5.02}_{-3.32}$ & \nodata\\
kT
(keV)&0.67$_{-0.03}^{+0.03}$&0.67 (t) & 0.67 (t) & 0.67 (t)\\
n$_{\rm e}$ (cm$^{-3}$)&$3.6\times10^{-2}$&$1.5\times10^{-2}$&$4.3\times10^{-3}$&$1.0\times10^{-3}$\\
O/O$_\odot$\tablenotemark{b}&0.8$^{+1.4}_{-0.7}$&0.6$^{+0.5}_{-0.3}$&1.0$^{+0.6}_{-0.4}$&1.3$^{+0.7}_{-0.6}$\\
Mg/Mg$_\odot$\tablenotemark{b}&1 (f)&0.6$^{+0.4}_{-0.3}$&0.8$^{+0.4}_{-0.3}$&0.5$^{+0.5}_{-0.4}$\\
Si/Si$_\odot$\tablenotemark{b}&6.0$^{+5.8}_{-2.9}$&1.2$^{+0.7}_{-0.5}$&1.2$^{+0.6}_{-0.5}$ & 3.4$^{+2.3}_{-1.8}$\\
Fe/Fe$_\odot$\tablenotemark{b}&0.4$^{+0.5}_{-0.2}$&0.3$^{+0.1}_{-0.1}$& 0.3$^{+0.1}_{-0.1}$ & 0.4$^{+0.2}_{-0.1}$\\
$\tau_{\rm u}$ (10$^{12}$
s~cm$^{-3}$)\tablenotemark{c}&4.1$^{+11.2}_{-2.1}$&4.1 (t) & 4.1 (t) &
4.1 (t)\\
f$_{\rm 0.5-2~keV}$ (10$^{-14}$~erg~s$^{-1}$~cm$^{-2}$)\tablenotemark{d}&0.57$_{-0.09}^{+0.03}$&1.39$_{-0.08}^{+0.05}$&2.01$_{-0.10}^{+0.08}$&2.06$_{-0.21}^{+0.11}$\\
f$_{\rm 2-10~keV}$ (10$^{-14}$~erg~s$^{-1}$~cm$^{-2}$)\tablenotemark{d}&3.70$_{-0.26}^{+0.26}$&2.39$_{-0.25}^{+0.23}$&2.04$_{-0.36}^{+0.34}$&0.10$_{-0.02}^{+0.10}$\\
L$_{\rm shock, 0.5-10
  keV}$(10$^{40}$~erg~s$^{-1}$)\tablenotemark{e}&1.26&5.26&8.61&8.86\\
L$^{\rm HMXB}_{\rm host}$ (10$^{40}$~erg~s$^{-1}$)\tablenotemark{e}&\nodata&\nodata&12.78&\nodata\\
L$_{\rm AGN, 0.5-2
  keV}$(10$^{40}$~erg~s$^{-1}$)\tablenotemark{e}&5.41&3.42&0.00&0.00\\
L$_{\rm AGN, 2-10
  keV}$(10$^{40}$~erg~s$^{-1}$)\tablenotemark{e}&15.58&9.84&0.00&0.00\\
\enddata

\tablenotetext{a}{ Best-fit model: $N_{\rm H,~Galactic} \times
  \{f_{\rm nuclear~NuSTAR} \times N_{\rm H,~nucleus} \times ({\rm
    MYTorus}[N_{\rm H,~AGN}, PL_{\rm AGN}] + f_{\rm C-thin} \times
  PL_{\rm AGN} + N_{\rm H,~nuclear~HMXB} \times PL^{\rm cutoff}_{\rm
    nuclear~HMXB}) + (N_{\rm H,~host~HMXB} \times PL^{\rm cutoff}_{\rm
    host~HMXB} + {\rm vpshock})\}$, where $N_{\rm H,~Galalactic}$ =
  1.26 $\times$ 10$^{20}$ cm$^{-2}$, the Galactic column density in
  the direction of Mrk~231 (Dickey \& Lockman 1990), $f_{\rm
    nuclear~NuSTAR}$ is the fraction ($< 1$) of the NuSTAR spectrum of
  Teng et al. (2014), minus the two MEKAL models of the nuclear
  diffuse emission, within the given aperture, $N_{\rm H,~nucleus}$ is
  an additional absorbing column in the line of sight toward the
  nucleus, not seen in the shallow {\em Chandra} spectrum of Teng et
  al. (2014), $N_{\rm H,~AGN}$ is the absorbing column of the AGN
  emission calculated as part of the MYTorus model and $PL_{\rm AGN}$
  is the direct AGN emission within the MYTorus model that also
  includes the scattered fraction and Fe lines, $f_{\rm C-thin}$ is
  the fraction (= 0.19) of the ``leaked'' direct AGN emission, $N_{\rm
    H,~nuclear~HMXB}$ $\times$ $PL^{\rm cutoff}_{\rm nuclear~HMXB}$ is
  the highly obscured emission from HMXBs in the nucleus, $N_{\rm
    H,~host~HMXB}$ $\times$ $PL^{\rm cutoff}_{\rm host~HMXB}$ is the
  emission from HMXBs outside of the nucleus (only detected in annulus
  \#3), and vpshock is a constant temperature, plane-parallel shock
  plasma model of the emission from the hot diffuse gas (see text for
  more detail). This best-fit model has $\chi^2$ of 227.4 for 236
  degrees of freedom ($\chi^2_\nu$=0.96). (t) indicates that the value
  of the parameter is tied to be the same in all regions.  (f)
  indicates that the value of the parameter is held fixed.}

\tablenotetext{b}{Atomic abundance relative to Solar value.  All other
elemental abundances fixed at Solar.}

\tablenotetext{c}{Upper limit on the ionization time scale ($\tau =
  \int{n_e t dt}$).  The lower limit is fixed at 0.}

\tablenotetext{d}{Observed flux.}

\tablenotetext{e}{Absorption corrected luminosity.}
\label{tab:shockfits}
\end{deluxetable}
%\end{turnpage}

\tablenum{3}
%\begin{turnpage}
\begin{deluxetable}{ccccc}
%\rotate
\tablecolumns{5}
\tabletypesize{\tiny}
\setlength{\tabcolsep}{0.01in}
\tablecaption{a) Simultaneous Model Fit to the Neutral Outflow Regions
  (MEKAL models)}
\tablewidth{0pt}
\tablehead{\colhead{Model} & \colhead{East} & \colhead{South} &\colhead{West}& \colhead{North}\\
\colhead{Parameter\tablenotemark{a}} &&&&
}
\startdata
Position Angle($^\circ$)&45--135&135--225&225--315&--45--45\\
Radial Distance (kpc) & 1.0--2.0 & 1.0--2.0&1.0--2.0&1.0--2.0\\
Hardness Ratio & $-0.167\pm0.048$&$-0.104\pm0.061$&$0.034\pm0.066$&$0.036\pm0.055$\\
Ratio to Nuclear NuSTAR Model & 0.0096$^{+0.0007}_{-0.0007}$&0.0096 (t)
&0.0096 (t) & 0.0096 (t)\\ 
N$_{\rm H, AGN}$
(10$^{24}$~cm$^{-2}$)&0.112 (f) & 0.112 (t) & 0.112
(t) & 0.112 (t)\\
C-thin Fraction&0.194 (f) & 0.194 (t) & 0.194 (t) &
0.194 (t)\\
kT (keV)&0.75$_{-0.14}^{+0.13}$ & 0.75 (t) & 0.75 (t) & 0.75 (t)\\
n$_{\rm e}$
(cm$^{-3}$)&$5.4\times10^{-2}$&$3.6\times10^{-2}$&$2.8\times10^{-2}$&$4.7\times10^{-2}$\\
O/O$_\odot$\tablenotemark{b}& 1 (f) & 1 (f) & 1 (f) & 1 (f)\\
Mg/Mg$_\odot$\tablenotemark{b}&1 (f) & 1 (f) & 1 (f) & 1 (f)\\
Si/Si$_\odot$\tablenotemark{b}&4.0$^{+3.7}_{-2.4}$&1 (f) & 1 (f) &
1 (f)\\
Fe/Fe$_\odot$\tablenotemark{b}&0.3$^{+0.3}_{-0.2}$&0.3 (f) & 0.3 (f) &
0.1$^{+0.4}_{-0.1}$\\
f$_{\rm 0.5-2~keV}$ (10$^{-15}$~erg~s$^{-1}$~cm$^{-2}$)\tablenotemark{c}&2.02$_{-0.27}^{+0.18}$&1.19$_{-0.16}^{+0.13}$&0.97$_{-0.15}^{+0.14}$&1.43$_{-0.26}^{+0.19}$\\
f$_{\rm 2-10~keV}$ (10$^{-15}$~erg~s$^{-1}$~cm$^{-2}$)\tablenotemark{c}&8.78$_{-0.65}^{+0.65}$&8.70$_{-0.64}^{+0.55}$&8.68$_{-0.65}^{+0.54}$&8.73$_{-0.63}^{+0.56}$\\
L$_{\rm MEKAL}$
(10$^{39}$~erg~s$^{-1}$)\tablenotemark{d}&6.51&2.57&1.54&3.79\\
L$_{\rm AGN, 0.5-2
  keV}$(10$^{40}$~erg~s$^{-1}$)\tablenotemark{d}&1.28&1.28&1.28&1.28\\
L$_{\rm AGN, 2-10
  keV}$(10$^{40}$~erg~s$^{-1}$)\tablenotemark{d}&3.68 & 3.68 & 3.68 & 3.68\\
\enddata

\tablenotetext{a}{Best-fit model: $N_{\rm H,~Galactic} \times
    \{f_{\rm nuclear~NuSTAR} \times N_{\rm H,~nucleus} \times ({\rm
      MYTorus}[N_{\rm H,~AGN}, PL_{\rm AGN}] + f_{\rm C-thin} \times
    PL_{\rm AGN} + N_{\rm H,~nuclear~HMXB} \times PL^{\rm cutoff}_{\rm
      nuclear~HMXB}) + {\rm MEKAL}_1\}$, where $N_{\rm H,~Galalactic}$
    = 1.26 $\times$ 10$^{20}$ cm$^{-2}$, the Galactic column density
    in the direction of Mrk~231 (Dickey \& Lockman 1990), $f_{\rm
      nuclear~NuSTAR}$ is the fraction ($< 1$) of the NuSTAR spectrum
    of Teng et al. (2014), minus the two MEKAL models of the nuclear
    diffuse emission, within the given aperture, $N_{\rm H,~nucleus}$
    is an additional absorbing column in the line of sight toward the
    nucleus, not seen in the shallow {\em Chandra} spectrum of Teng et
    al. (2014), $N_{\rm H,~AGN}$ is the absorbing column of the AGN
    emission calculated as part of the MYTorus model and $PL_{\rm
      AGN}$ is the direct AGN emission within the MYTorus model that
    also includes the scattered fraction and Fe lines, $f_{\rm
      C-thin}$ is the fraction (= 0.19) of the ``leaked'' direct AGN
    emission, $N_{\rm H,~nuclear~HMXB}$ $\times$ $PL^{\rm cutoff}_{\rm
      nuclear~HMXB}$ is the highly obscured emission from HMXBs in the
    nucleus, and MEKAL$_1$ is a MEKAL model of the emission from the
    hot diffuse gas (see text for more detail). This best-fit model
    has $\chi^2$ of 87.2 for 72 degrees of freedom
    ($\chi^2_\nu$=1.21). (t) indicates that the value of the parameter
    is tied to be the same in all regions.  (f) indicates that the
    value of the parameter is held fixed. }

\tablenotetext{b}{Atomic abundance relative to Solar value.  All other
  elemental abundances fixed at Solar.}  

\tablenotetext{c}{Observed flux.}

\tablenotetext{d}{Absorption corrected luminosity.}

\label{tab:wind}
\end{deluxetable}
%\end{turnpage}

\tablenum{3}
%\begin{turnpage}
\begin{deluxetable}{ccccc}
%\rotate
\tablecolumns{5}
\tabletypesize{\tiny}
\setlength{\tabcolsep}{0.01in}
\tablecaption{b) Simultaneous Model Fit to the Neutral Outflow Regions
  (shock models)}
\tablewidth{0pt}
\tablehead{\colhead{Model} & \colhead{East} & \colhead{South} &\colhead{West}& \colhead{North}\\
\colhead{Parameter\tablenotemark{a}} &&&&
}
\startdata
Position Angle($^\circ$)&45--135&135--225&225--315&--45--45\\
Radial Distance (kpc) & 1.0--2.0 & 1.0--2.0&1.0--2.0&1.0--2.0\\
Hardness Ratio & $-0.167\pm0.048$&$-0.104\pm0.061$&$0.034\pm0.066$&$0.036\pm0.055$\\
Ratio to Nuclear NuSTAR Model & 0.0097$^{+0.0007}_{-0.0007}$&0.0097 (t)
&0.0097 (t) & 0.0097 (t)\\ 
N$_{\rm H, AGN}$
(10$^{24}$~cm$^{-2}$)&0.112 (f) & 0.112 (t) & 0.112
(t) & 0.112 (t)\\
C-thin Fraction&0.194 (f) & 0.194 (t) & 0.194 (t) &
0.194 (t)\\
kT (keV)&0.70$_{-0.14}^{+0.17}$ & 0.70 (t) & 0.70 (t) & 0.70 (t)\\
n$_{\rm e}$
(cm$^{-3}$)&$5.7\times10^{-2}$&$3.2\times10^{-2}$&$2.8\times10^{-2}$&$4.5\times10^{-2}$\\
O/O$_\odot$\tablenotemark{b}& 0.8$^{+2.8}_{-0.8}$ & 1 (f) & 1 (f) & 1 (f)\\
Mg/Mg$_\odot$\tablenotemark{b}&1 (f) & 1 (f) & 1 (f) & 1 (f)\\
Si/Si$_\odot$\tablenotemark{b}&3.9$^{+4.0}_{-2.2}$& 1 (f) & 1 (f)
& 4.3$^{+8.2}_{-3.6}$\\
Fe/Fe$_\odot$\tablenotemark{b}&0.2$^{+0.4}_{-0.1}$&0.4$^{+3.9}_{-0.3}$ & 0.3 (f) &
0.1$^{+0.6}_{-0.1}$\\
$\tau_{\rm u}$ (10$^{12}$
s~cm$^{-3}$)\tablenotemark{c}&$> 2.8$ & $> 2.8$ (t) & $> 2.8$ (t) & $>
2.8$ (t)\\
f$_{\rm 0.5-2~keV}$ (10$^{-15}$~erg~s$^{-1}$~cm$^{-2}$)\tablenotemark{d}&2.00$_{-0.38}^{+0.16}$&1.17$_{-0.37}^{+0.12}$&0.98$_{-0.14}^{+0.14}$&1.46$_{-0.33}^{+0.17}$\\
f$_{\rm 2-10~keV}$
(10$^{-15}$~erg~s$^{-1}$~cm$^{-2}$)\tablenotemark{d}&8.80$_{-0.62}^{+0.61}$&8.73$_{-0.62}^{+0.62}$&8.72$_{-0.62}^{+0.63}$&8.77$_{-0.62}^{+0.62}$\\
L$_{\rm shock, 0.5-10
  keV}$(10$^{39}$~erg~s$^{-1}$)\tablenotemark{e}&6.29&2.40&1.53&3.82\\
L$_{\rm AGN, 0.5-2
  keV}$(10$^{40}$~erg~s$^{-1}$)\tablenotemark{e}&1.28&1.28&1.28&1.28\\
L$_{\rm AGN, 2-10
  keV}$(10$^{40}$~erg~s$^{-1}$)\tablenotemark{e}&3.70 & 3.70 & 3.70 & 3.70\\
\enddata

\tablenotetext{a}{ Best-fit model: $N_{\rm H,~Galactic} \times
  \{f_{\rm nuclear~NuSTAR} \times N_{\rm H,~nucleus} \times ({\rm
    MYTorus}[N_{\rm H,~AGN}, PL_{\rm AGN}] + f_{\rm C-thin} \times
  PL_{\rm AGN} + N_{\rm H,~nuclear~HMXB} \times PL^{\rm cutoff}_{\rm
    nuclear~HMXB}) + {\rm vpshock}\}$, where $N_{\rm H,~Galalactic}$ =
  1.26 $\times$ 10$^{20}$ cm$^{-2}$, the Galactic column density in
  the direction of Mrk~231 (Dickey \& Lockman 1990), $f_{\rm
    nuclear~NuSTAR}$ is the fraction ($< 1$) of the NuSTAR spectrum of
  Teng et al. (2014), minus the two MEKAL models of the nuclear
  diffuse emission, within the given aperture, $N_{\rm H,~nucleus}$ is
  an additional absorbing column in the line of sight toward the
  nucleus, not seen in the shallow {\em Chandra} spectrum of Teng et
  al. (2014), $N_{\rm H,~AGN}$ is the absorbing column of the AGN
  emission calculated as part of the MYTorus model and $PL_{\rm AGN}$
  is the direct AGN emission within the MYTorus model that also
  includes the scattered fraction and Fe lines, $f_{\rm C-thin}$ is
  the fraction (= 0.19) of the ``leaked'' direct AGN emission, $N_{\rm
    H,~nuclear~HMXB}$ $\times$ $PL^{\rm cutoff}_{\rm nuclear~HMXB}$ is
  the highly obscured emission from HMXBs in the nucleus, and vpshock
  is a constant temperature, plane-parallel shock plasma model of the
  emission from the hot diffuse gas (see text for more detail). This
  best-fit model has $\chi^2$ of 86.3 for 68 degrees of freedom
  ($\chi^2_\nu$=1.27). (t) indicates that the value of the parameter
  is tied to be the same in all regions.  (f) indicates that the value
  of the parameter is held fixed.}

\tablenotetext{b}{Atomic abundance relative to Solar value.  All other
  elemental abundances fixed at Solar.}

\tablenotetext{c}{Upper limit on the ionization time scale ($\tau =
  \int{n_e t dt}$).  The lower
limit is fixed at 0.}

\tablenotetext{d}{Observed flux.}

\tablenotetext{e}{Absorption corrected luminosity.}
\label{tab:wind}
\end{deluxetable}
%\end{turnpage}

\tablenum{4}
%\begin{turnpage}
\begin{deluxetable}{ccccc}
%\rotate
\tablecolumns{5}
\tabletypesize{\tiny}
\setlength{\tabcolsep}{0.01in}
\tablecaption{Simultaneous Model Fit to the Tidal Arc Regions}
\tablewidth{0pt}
\tablehead{\colhead{Model} & \colhead{Arc} & \colhead{East} &\colhead{West}& \colhead{North}\\
\colhead{Parameter\tablenotemark{a}} &&&&
}
\startdata
Position Angle($^\circ$)&125--215&35--125&215--305&--55--35\\
Radial Distance (kpc) & 2.0--4.5 & 2.0--4.5 & 2.0--4.5 & 2.0--4.5\\
Hardness Ratio & $-0.708^{+0.065}_{-0.061}$&$-0.613\pm0.062$&$-0.514^{+0.080}_{-0.074}$&$-0.710^{+0.064}_{-0.060}$\\
Ratio to Nuclear NuSTAR Model & 0.0049$^{+0.0006}_{-0.0006}$&0.0049 (t)&0.0049 (t)&0.0049 (t)\\
N$_{\rm H, AGN}$
(10$^{24}$~cm$^{-2}$) &0.112 (f) &0.112 (t)&0.112
(t)&0.112 (t)\\
C-thin Fraction&0.194 (f)&0.194 (t) &0.194 (t)&0.194 (t)\\
kT (keV)&0.65$_{-0.07}^{+0.06}$&0.65 (t) & 0.65 (t) & 0.65 (t)\\
n$_{\rm e}$ (cm$^{-3}$)&$1.6\times10^{-2}$&$2.2\times10^{-2}$&$1.5\times10^{-2}$&$2.3\times10^{-2}$\\
O/O$_\odot$\tablenotemark{b}&2.4$^{+31.4}_{-1.7}$&0.6$^{+1.4}_{-0.5}$
& 1.2$^{+6.9}_{-1.2}$ & 0.3$^{+1.1}_{-0.3}$\\
Mg/Mg$_\odot$\tablenotemark{b}&1.0$^{+17.4}_{-1.0}$&0.7$^{+1.3}_{-0.7}$ & 0.9$^{+6.2}_{-0.9}$ &1.2$^{+1.7}_{-0.9}$ \\
Si/Si$_\odot$\tablenotemark{b}&3.7$^{+33.5}_{-2.7}$&0.9$^{+1.6}_{-0.9}$ & 1.6$^{+7.9}_{-1.6}$ &0.3$^{+1.3}_{-0.3}$ \\
Fe/Fe$_\odot$\tablenotemark{b}&0.6$^{+8.3}_{-0.4}$&0.2$^{+0.3}_{-0.1}$ & 0.3$^{+1.7}_{-0.2}$  &0.3$^{+0.4}_{-0.2}$ \\
E$_{\rm line, 1}$ (keV)&1.242$_{-0.045}^{+0.028}$&\nodata&\nodata&\nodata\\
EW$_{\rm line, 1}$ (keV)&0.163$^{+0.179}_{-0.101}$&\nodata&\nodata&\nodata\\
Line ID& Ne X or Fe XIX&\nodata&\nodata&\nodata\\
E$_{\rm line, 2}$ (keV)&1.670$_{-0.060}^{+0.157}$&\nodata&\nodata&\nodata\\
EW$_{\rm line, 2}$ (keV)&0.281$_{-0.221}^{+0.245}$&\nodata&\nodata&\nodata\\
Line ID&Si XIII&\nodata&\nodata&\nodata\\
f$_{\rm 0.5-2~keV}$ (10$^{-15}$~erg~s$^{-1}$~cm$^{-2}$)\tablenotemark{c}&2.92$_{-1.05}^{+0.21}$&2.51$_{-0.56}^{+0.17}$&1.53$_{-0.57}^{+0.13}$&2.86$_{-0.63}^{+0.21}$\\
f$_{\rm 2-10~keV}$ (10$^{-15}$~erg~s$^{-1}$~cm$^{-2}$)\tablenotemark{c}&4.46$_{-0.48}^{+0.47}$&4.48$_{-0.49}^{+0.50}$&4.42$_{-0.43}^{+0.50}$&4.49$_{-0.51}^{+0.50}$\\
L$_{\rm MEKAL}$
(10$^{40}$~erg~s$^{-1}$)\tablenotemark{d}&1.06&1.02&0.56&1.17\\
L$_{\rm AGN, 0.5-2
  keV}$(10$^{40}$~erg~s$^{-1}$)\tablenotemark{d}&0.64&0.64&0.64&0.64\\
L$_{\rm AGN, 2-10
  keV}$(10$^{40}$~erg~s$^{-1}$)\tablenotemark{d}&1.85&1.85&1.85&1.85\\
\enddata

\tablenotetext{a}{Best-fit model: $N_{\rm H,~Galactic} \times \{f_{\rm
    nuclear~NuSTAR} \times N_{\rm H,~nucleus} \times ({\rm
    MYTorus}[N_{\rm H,~AGN}, PL_{\rm AGN}] + f_{\rm C-thin} \times
  PL_{\rm AGN} + N_{\rm H,~nuclear~HMXB} \times PL^{\rm cutoff}_{\rm
    nuclear~HMXB}) + {\rm line[1-2]} + {\rm MEKAL}_1\}$, where $N_{\rm
    H,~Galalactic}$ = 1.26 $\times$ 10$^{20}$ cm$^{-2}$, the Galactic
  column density in the direction of Mrk~231 (Dickey \& Lockman 1990),
  $f_{\rm nuclear~NuSTAR}$ is the fraction ($< 1$) of the NuSTAR
  spectrum of Teng et al. (2014), minus the two MEKAL models of the
  nuclear diffuse emission, within the given aperture, $N_{\rm
    H,~nucleus}$ is an additional absorbing column in the line of
  sight toward the nucleus, not seen in the shallow {\em Chandra}
  spectrum of Teng et al. (2014), $N_{\rm H,~AGN}$ is the absorbing
  column of the AGN emission calculated as part of the MYTorus model
  and $PL_{\rm AGN}$ is the direct AGN emission within the MYTorus
  model that also includes the scattered fraction and Fe lines,
  $f_{\rm C-thin}$ is the fraction (= 0.19) of the ``leaked'' direct
  AGN emission, $N_{\rm H,~nuclear~HMXB}$ $\times$ $PL^{\rm
    cutoff}_{\rm nuclear~HMXB}$ is the highly obscured emission from
  HMXBs in the nucleus, line[1 -- 2] are Gaussian fits to the emission
  lines, and MEKAL$_1$ is a MEKAL model of the emission from the hot
  diffuse gas (see text for more detail).  This best-fit model has
  $\chi^2$ of 49.1 for 67 degrees of freedom ($\chi^2_\nu$=0.73). (t)
  indicates that the value of the parameter is tied to be the same in
  all regions.  (f) indicates that the value of the parameter is held
  fixed.}

\tablenotetext{b}{Atomic abundance relative to Solar value.  All other
  elemental abundances fixed at Solar.}

\tablenotetext{c}{Observed flux.}

\tablenotetext{d}{Absorption corrected luminosity.}

\label{tab:arcfits}
\end{deluxetable}
%\end{turnpage}

\clearpage

\begin{figure}
%1
%\epsscale{1}
\plotone{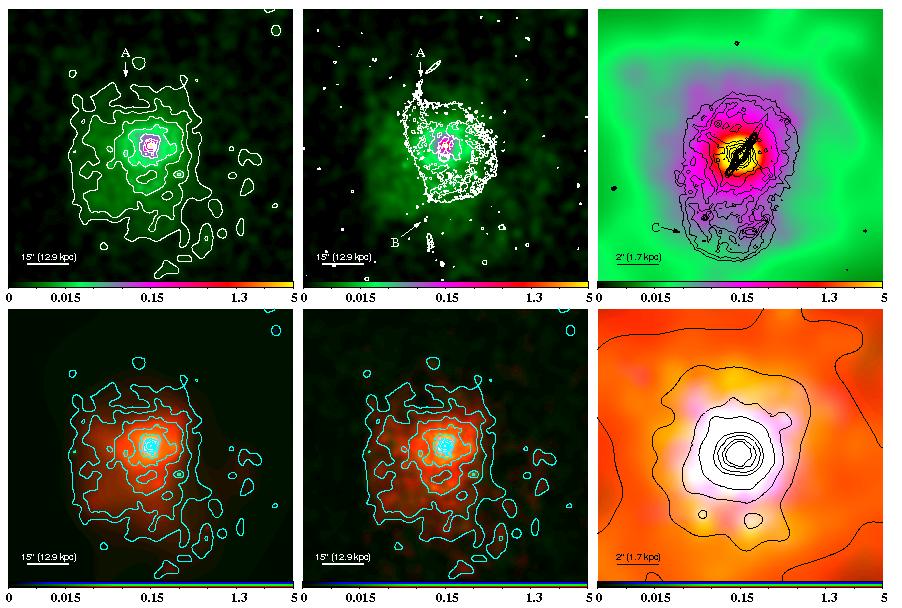}
\caption{ ({\em Top}) Adaptively smoothed full X-ray band (0.5 -– 8
  keV) merged images from all observations (0.52 Msec in total), on a
  logarithmic scale. North is up and East is to the left. The overlaid
  contours are ({\em left}) the full 0.5 -– 8 keV contours at 2, 3, 5,
  10, 25, 50, 100, 500, 1000, and 5000 $\sigma$, ({\em middle}) the
  contours of the optical {\em HST} image from RV11, and ({\em right})
  the contours of the {\em HST} optical image zoomed-in to show the
  tidal star-forming arc $\sim$3.5 kpc south of the nucleus. The
  cross-like pattern in the center of the optical image in both the
  middle and right panels is an artifact of the strong central
  PSF. Labels A, B, and C point to the northern tidal tail, the
  southern tidal tail, and the southern tidal arc, respectively. ({\em
    Bottom}) False-color X-ray images of adatively smoothed images
  overlaid with the full X-ray band contours from the upper left
  panel.  In all three of these bottom panels, red represents the 0.5
  -- 1 keV emission, green is 1 -- 2 keV, and blue is 2 -- 8 keV. The
  bottom left panel is smoothed using {\em csmooth} to emphasize the
  large-scale structure. The middle panel is smoothed using the {\em
    imageadapt} script designed to better show the small-scale
  structures. The bottom right panel is a zoomed-in version of the
  middle panel.}
\label{fig:mrk231_panels}
\end{figure}

\begin{figure*}[ht]
%2
%\epsscale{.3}
\centering
\includegraphics[width=5in]{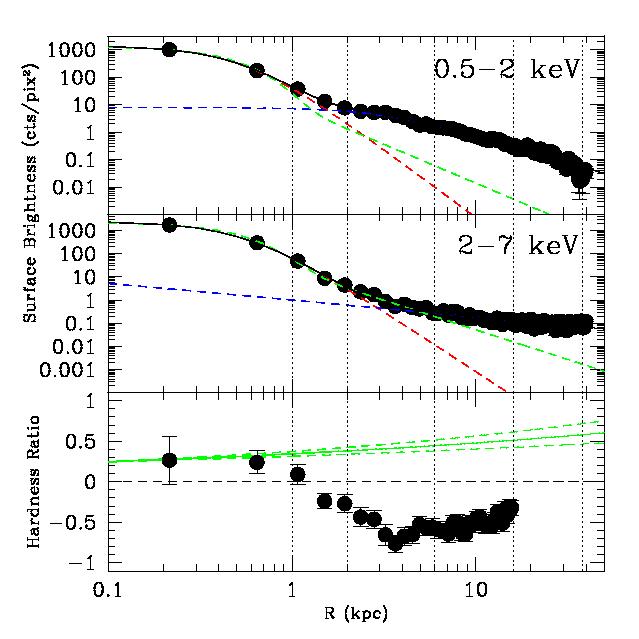}
\caption{Radial profiles of Mrk 231 in the 0.5--2~keV and 2--7~keV
  bands.  In the top two panels, the background-subtracted data points
  are represented by filled circles.  A two-component $\beta$ model is
  fitted to each profile (see text and Table~1) with the best fit
  shown as a black curve.  The red and blue dash lines represent the
  inner and outer components of each model, respectively.  The green
  dash lines in the top two panels represent the simulated PSF fit
  using two-component $\beta$ models. The soft emission is extended
  beyond $\sim$1~kpc and the hard emission is extended beyond
  $\sim$6~kpc.  The bottom panel is a radial profile of the hardness
  ratio, emphasizing the softness of the extended emission.  The green
  curve shows the hardness ratio of the simulated PSF for comparison.
  The vertical dotted lines show the division between the annuli
  extraction regions for Tables 2a.and 2b The drop in the hardness
  ratio from $\sim$1 kpc to $\sim$6 kpc reflects the increasing
  importance of the extended soft X-ray component relative to that of
  the central quasar. Beyond 16 kpc, the hardness ratio profile is not
  shown because it is unreliable -- there are very few counts in the
  hard X-ray band on this scale and the soft X-ray emission is also
  very patchy.}
\label{fig:rprofile}
\end{figure*}

\begin{figure*}[ht]
%3
%\epsscale{.3}
\centering
\includegraphics[width=6.5in]{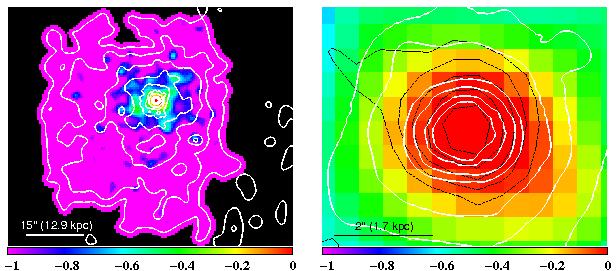}
\caption{Pixel-to-pixel hardness ratio map of Mrk~231.  North is up
  and East is to the left. In both panels, the image is smoothed with
  a 4-pixel Gaussian and the {\it Chandra} 0.5--8~keV contours are
  shown in white for comparison.  The left panel shows the extended
  soft emission and the right panel zooms in on the nuclear region.
  In the right panel, the contours of the PSF hardness ratio (Figure
  2) are shown as dashed black lines for comparison.  Note that since
  the PSF has a different hardness ratio than the data, this is only a
  comparison of the morphology of the hardness ratio, not its absolute
  value.  The faint northeast tail in the PSF contours is an artifact
  of the simulated PSF. Note the slight asymmetry of the nuclear
  hardness ratio map (color map in the right panel) extending to the
  west.}
\label{fig:hr}
\end{figure*}

\begin{figure}
%4
%\epsscale{1}
\plotone{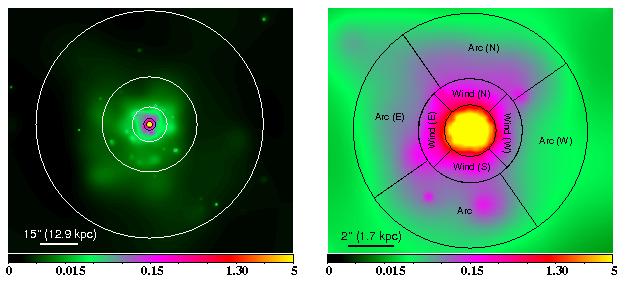}
\caption{Full X-ray band image showing the different extraction
  regions used for the spectral analysis in \S 3.2. North is up and East 
  is to the left. 
  ({\em Left}) Annular regions.  The nucleus and annuli \#1-4 have
  outer radii $R$ = 1.0 kpc, 2.0 kpc, 6 kpc, 16 kpc, and 40 kpc,
  respectively. ({\em Right}) The outflow and tidal arc extraction
  regions. The outflow extraction regions corresponds to annulus \#1 (1 --
  2 kpc) divided into four equal-size quadrants covering PA = --45 --
  +45$^\circ$ (N), +45 -- +135 (E), +135 -- +225 (S), and +225 --
  +315$^\circ$ (W). The tidal arc region extends over $R$ = 2.0 -- 4.5
  kpc and PA = +125 -- +215$^\circ$ (Arc). The comparison regions span
  PA = +35 -- +125$^\circ$ (E), +215 -- +305$^\circ$ (W), --55 --
  +35$^\circ$ (N). }
\label{fig:extract}
\end{figure}

\begin{figure*}[ht]
%5
%\epsscale{.3}
\centering
\includegraphics[width=4in, angle=270]{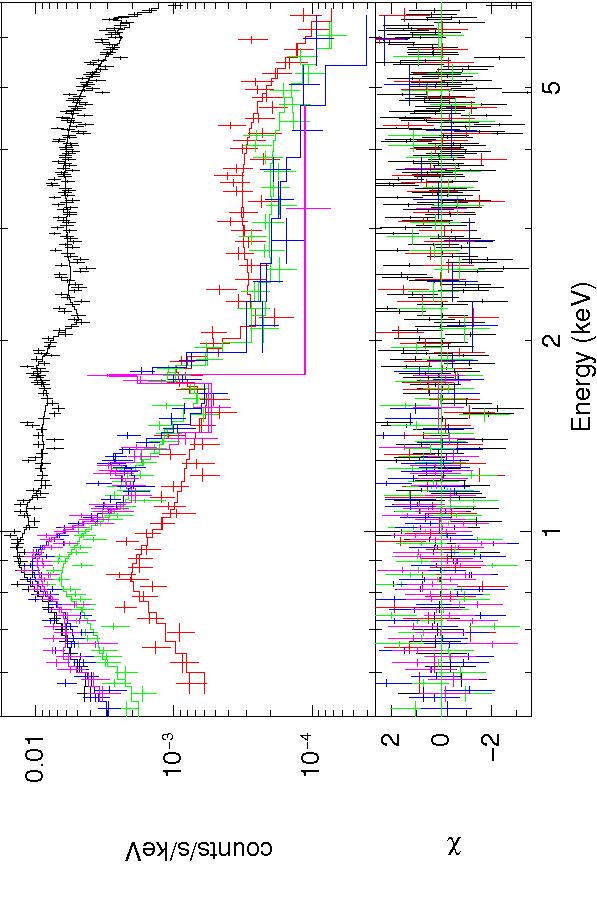}
\caption{({\em Top}) Spectra extracted from the annular regions shown
  with the best-fit model to the simultaneous fit. ({\em Bottom})
  Residuals (data minus model) where error bars of size one correspond
  to one sigma. The colors represent the nucleus binned to 8$\sigma$
  above the background (black), annulus \#1 binned to 8$\sigma$ above
  the background (red), annulus \#2 binned to 4$\sigma$ above the
  background (green), annulus \#3 binned to 4$\sigma$ above the
  background (blue), and annulus \#4 binned to 3$\sigma$ above the
  background (magenta).  Note the presence of Mg XI 1.352 keV and Si
  XIII 1.864 keV in many of these spectra. The model to annulus \#3 is
  underestimated near 6~keV.  This may indicate the presence of Fe~XXV
  6.7 keV outside of the nucleus (see Figure~11 for a more detailed
  fitting of the spectrum).  The parameters of the best fits are
  listed in Table 2a.  }
\label{fig:allregions}
\end{figure*}

\begin{figure*}[ht]
%6
%\epsscale{.3}
\centering
\includegraphics[width=4in, angle=270]{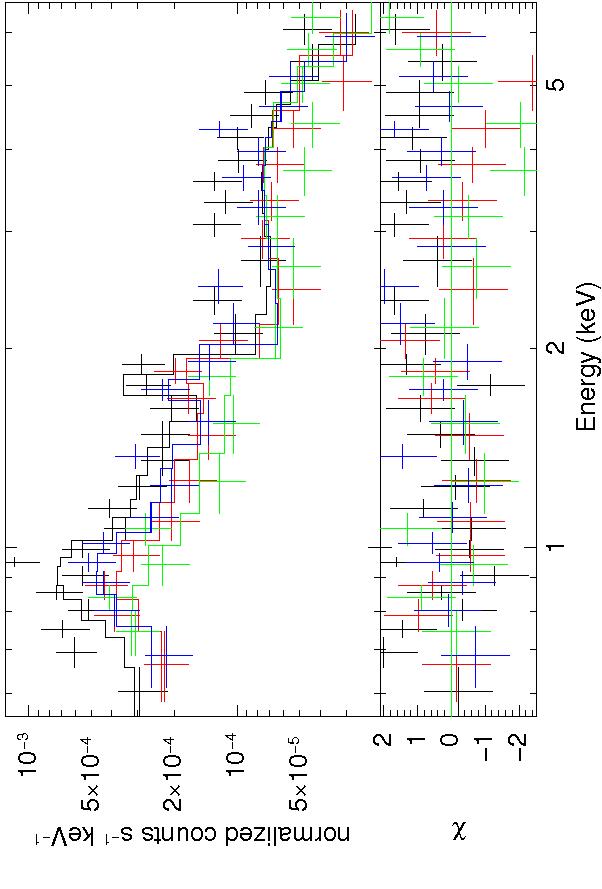}
\caption{({\em Top}) Spectra extracted from the outflow regions shown
  with the best-fit model to the simultaneous fit.  ({\em Bottom})
  Residuals (data minus model) where error bars of size one correspond
  to one sigma. Here, black is east, red is south, green is west, blue is
  north. The spectra are binned to at least 15 counts per bin.  The
  best-fit parameters are listed in Table 3a. Si XIII 1.864 keV is
  detected in the eastern and northern regions. }
\label{fig:wind}
\end{figure*}

\begin{figure*}[ht]
%7
%\epsscale{.3}
\centering
\includegraphics[width=4in, angle=270]{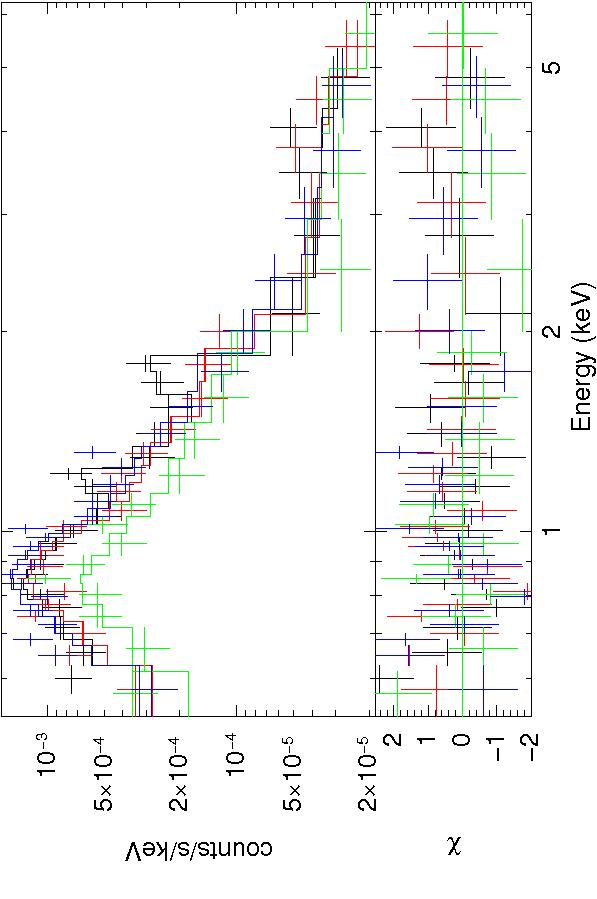}
\caption{({\em Top}) Spectra extracted from the tidal arc regions
  shown with the best-fit model to the simultaneous fit. ({\em
    Bottom}) Residuals (data minus model) where error bars of size one
  correspond to one sigma. The spectra are binned to at least 15
  counts per bin.  The black spectrum is the bright arc itself (PA:
  125 -- 215$^\circ$).  The red is the eastern comparison region (PA:
  35 -- 125$^\circ$), the green is the western comparison region
  (215 -- 305$^\circ$), and the blue is the northern comparison region
  (PA: --55 -- 35$^\circ$).  The parameters of the best fits are listed
  in Table 4.  Note the detection of the Mg XI 1.352 keV and Si XIII
  1.864 keV emission lines in the tidal arc spectrum, indicating the
  presence of $\alpha$-element enhancement due to the starburst in
  this region. }
\label{fig:arcs}
\end{figure*}

\begin{figure*}[ht]
%8
%\epsscale{.3}
\centering
\includegraphics[width=5in]{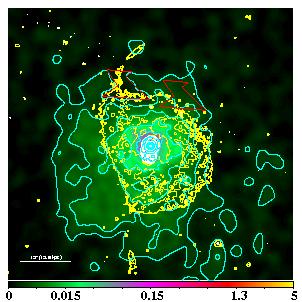}
\caption{ The polygonal extraction regions (in red) used to derive the
  X-ray spectra of the northern tidal tail (on the left) and a
  comparison region (on the right), shown against the contours of the
  full X-ray band image (cyan) and the {\em HST} optical image on
  large scale (yellow) and small scale (white). North is up and East
  is to the left. }
\label{fig:tail}
\end{figure*}

\clearpage

\begin{figure*}[ht]
%9
%\epsscale{.3}
\centering
\includegraphics[width=6.5in]{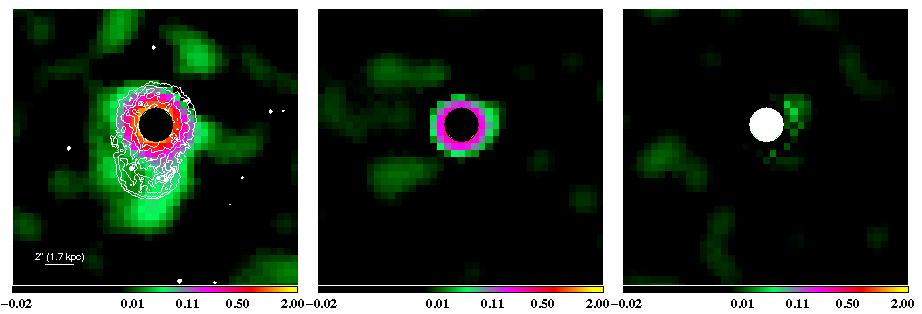}
\caption{Continuum-subtracted maps of the line emission in
  Mrk~231. North is up and East is to the left.  ({\em left}) Si XIII
  (1.70--1.95~keV) compared with the {\em HST} optical image
  (contours), ({\em middle}) Fe~K$\alpha$ (6.31--6.56~keV), and ({\em
    right}) Fe~XXV + Fe~XXVI (6.56--6.90~keV).  The emission is
  smoothed with a 4-pixel Gaussian.  The bright continuum emission
  from the central quasar makes the maps unreliable within the central
  3.0 kpc diameter region; this region is masked in the three
  panels. Filamentary Si XIII emission is seen out to $\sim$5
  kpc. Extended Fe XXV + Fe XXVI emission may also be present $\sim$3
  kpc north-west of the nucleus. }
\label{fig:linemaps}
\end{figure*}

\begin{figure*}[ht]
%10
%\epsscale{.3}
\centering
\includegraphics[width=4in, angle=270]{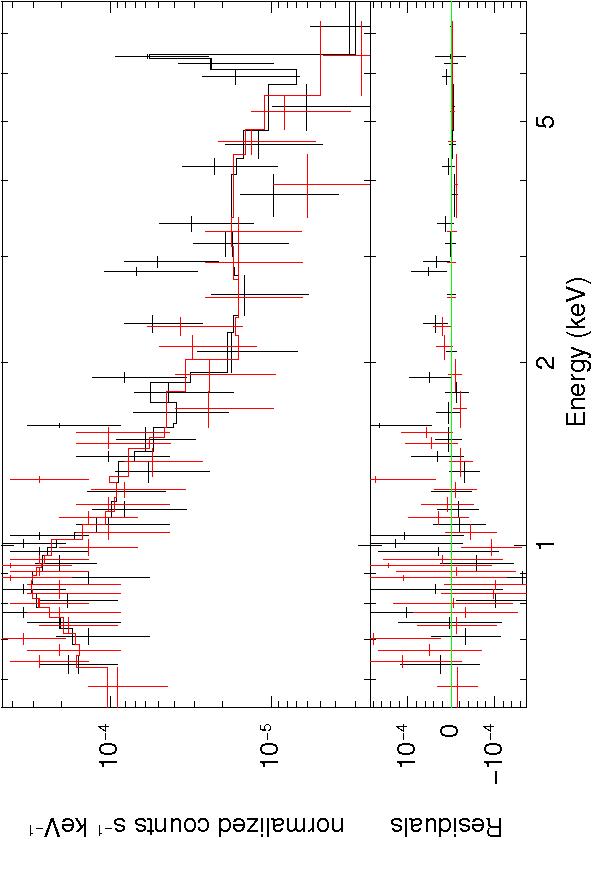}
\caption{ ({\em Top}) Spectrum (black data points) extracted from a
  2-arcsec diameter region centered on the brightest part of the Fe
  XXV 6.7 keV + Fe XXVI 6.9 keV extension north-west of the nucleus in
  Figure 9. ({\em Bottom}) Residuals (data minus model). The spectrum
  in red is from an equal-size region on the opposite side of the
  nucleus where there is no obvious Fe XXV + Fe XXVI emission in the
  line emission map.  The displayed spectra are binned to at least 3
  counts per bin, while the model was fit to unbinned data using Cash
  statistics. There is a tentative detection of Fe XXV 6.7 keV outside
  of the nucleus (see \S 3.2.5). }
\label{fig:linemaps}
\end{figure*}

\begin{figure*}[ht]
%11
%\epsscale{.3}
\centering
\includegraphics[width=4in, angle=270]{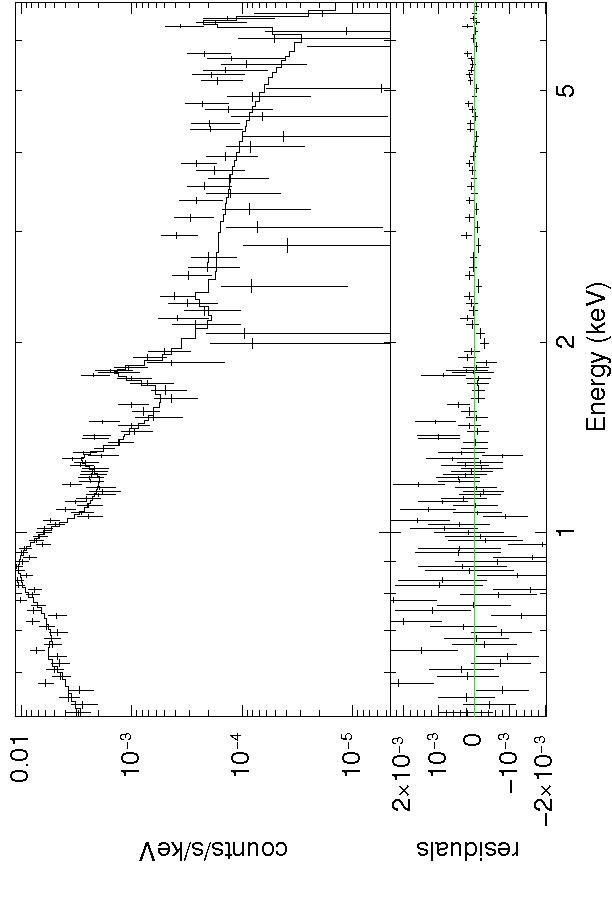}
\caption{({\em Top}) Spectrum of annulus \#3 with a finer binning than
  Figure~5 (at least 15 counts per bin).  ({\em Bottom}) Residuals
  (data minus model) where error bars of size one correspond to one
  sigma. The best-fit model shown in Figure~5 is applied with the
  addition of a narrow line at 6.7~keV to simulate Fe XXV emission. }
\label{fig:an3line}
\end{figure*}

\end{document}